\documentclass[11pt,a4]{article}
\pdfoutput=1
\usepackage{jheppub}
\usepackage{slashed}
\usepackage{graphicx,amssymb,amsmath,amsfonts}
\usepackage{hyperref}
\usepackage[utf8]{inputenc}

\def\be{\begin{equation}}
\def\ee{\end{equation}}

\def\bmat{\begin{pmatrix}}
\def\emat{\end{pmatrix}}

\def\bdet{\begin{vmatrix}}
\def\edet{\end{vmatrix}}

\DeclareMathOperator{\tr}{tr}

\numberwithin{equation}{section}

\def\bea{\begin{eqnarray}}
\def\eea{\end{eqnarray}}

\newcommand{\dprime}{{\prime\prime}}

\newcommand{\alp}{\ensuremath{\alpha^\prime}}

\newcommand{\sigf}{\sigma\!{}_f}

\newcommand{\pa}{\partial}
\newcommand{\w}{\omega}

\newcommand{\pil}{\frac\pi\ell}

\title{On the quantization of folded strings in non-critical dimensions}

\author[a]{Jacob Sonnenschein}
\author[b]{Dorin Weissman}

\affiliation[a]{The Raymond and Beverly Sackler School of Physics and Astronomy, Tel Aviv University, \\ Ramat Aviv 69978, Tel Aviv, Israel}
\affiliation[b]{Okinawa Institute of Science and Technology, \\ 
1919-1 Tancha, Onna-son, Okinawa 904-0495, Japan}

\emailAdd{cobi@post.tau.ac.il}
\emailAdd{dorin.weissman@oist.jp}

\abstract{Classical rotating closed string are folded strings. At the folding points the scalar curvature associated with the induced metric diverges. As a consequence one cannot properly quantize the fluctuations around the classical solution since there is no complete set of normalizable eigenmodes. Furthermore in the non-critical effective string action of Polchinski and Strominger, there is a divergence associated with the folds. We overcome this obstacle by putting a massive particle at each folding point which can be used as a regulator. Using this method we compute the spectrum of quantum fluctuations around the rotating string and the intercept of the leading Regge trajectory. The results we find are that the intercepts are $a=1$ and $a=2$ for the open and closed string respectively, independent of the target space dimension. We argue that in generic theories with an effective string description, one can expect corrections from finite masses associated with either the endpoints of an open string or the folding points on a closed string. We compute explicitly the corrections in the presence of these masses.}

\arxivnumber{2006.14634}

\begin{document}

\maketitle
\tableofcontents
\flushbottom
 
\section{Introduction}
The quantization of strings in non-critical dimensions and in particular in four spacetime dimensions is an utmost important question for the description of nature in terms of a string theory. An important step toward this goal has been achieved in \cite{Hellerman:2013kba}, where by using the Polchinski-Strominger (PS) effective action\cite{Polchinski:1991ax}, the rotating open strings at any $D$ dimensions and closed strings in $D\geq 5$ were quantized. The former case was shown to  admit an intercept $a=1$ like the one in critical dimensions, this being the leading order result for large angular momentum. However, the interesting case of rotating closed strings in four dimensions was not resolved in that paper. In the latter case there is a singularity in the PS action that could not be handled using the tools of that paper. The reason for that is that the classical rotating closed string develops two folding points, and the PS action diverges at those points. Folds occur also for bosonic strings in the critical dimension and, as we will show, form obstacles to the semiclassical quantization of the system. Thus the goal of this note is to provide a procedure of quantizing fluctuations around rotating strings in general, and in particular to tackle the quantization of closed strings in four dimensions.

In fact folded strings are quite generic configurations in string theory, so that proper quantization of them is a crucial task. Folded strings have been analyzed in various different circumstances, including: the 2D string theory dual of YM$_2$ \cite{Gross:1993hu}, 2D strings duals of certain lattice models\cite{Ganor:1994rm}, strings associated with QCD \cite{Pawelczyk:1994ex} and strong coupling \cite{Georgiou:2010an}, strings in curved spacetime \cite{Bars:1994xi}, strings falling into a black hole \cite{Bars:1994sv}, the structure inside the horizon of a black hole  \cite{Attali:2018goq,Giveon:2019gfk}, folded non-critical strings \cite{Maldacena:2005hi}, rotating folded strings in $AdS_5\times S^5$ (see \cite{Giombi:2010bj} and many references therein), glueballs as rotating folded strings \cite{Sonnenschein:2015zaa}, rotating open strings in magnetic fields \cite{Sonnenschein:2019bca}, and more \cite{Ando:2010nm}.

Folded maps from a worldsheet to a target space are characterized by the fact that the scalar curvature associated with the induced metric diverges on the folding points. For rotating strings the folding points move at the speed of light and these properties are related. It turns out, as was observed in \cite{Sonnenschein:2019bca}, that a folding point is a severe obstacle to quantizing the string. Because of the divergence at the folding points the eigenmodes of fluctuations of the string around the classical solution are not normalizable. Thus, the canonical quantization of the fluctuations around a folded string is ill defined.

There is no apparent problem in canonically quantizing in the standard way a closed string that carries angular momentum. The problem arises only when one uses the semiclassical procedure of quantizing the fluctuations around a classical folded rotating string solution. There are string systems where the latter procedure is the most natural one, especially in the effective string theory description of long strings \cite{Polchinski:1991ax,Aharony:2010cx,Aharony:2013ipa,Dubovsky:2012sh,Hellerman:2014cba}. Some specific examples include the stringy models of hadrons \cite{Sonnenschein:2016pim}  as well as certain other cases for instance \cite{Gubser:2002tv,Maldacena:2005hi,Attali:2018goq}.

A semiclassical description of fluctuations around rotating strings in particular is important due to the fact that rotating strings have asymptotically linear Regge trajectories,
\be J = \alp M^2 + \ldots \ee
or in the language of Regge theory \(\alpha(t) = \alp t + \ldots\). In \cite{Caron-Huot:2016icg} it was shown by imposing various consistency conditions on \(2\to2\) scattering amplitudes that any theory with weakly interacting massive higher spin particles is constrained to have linear trajectories at large \(t\), and the scattering amplitude \(A(s,t)\) in the limit of large positive \(s\) and \(t\) must coincide with the Veneziano amplitude. String theory at weak coupling and large \(N\) YM are the most prominent examples of theories to which this applies, but there is a large space of QFTs which have an effective string limit, and then the expansion around rotating strings is the most natural way to describe them. In \cite{Sever:2017ylk}, a followup work to \cite{Caron-Huot:2016icg}, it was found using the same methods that corrections corresponding to massive particles on the string
\be J = \alp M^2 + c_m \alp m^{3/2}M^{1/2} + \ldots \label{eq:trajgeneral} \ee
are also universal for the same space of theories, in the sense that it is the only possible term that still increases with the energy. Here \(c_m\) is an \(\mathcal O(1)\) dimensionless coefficient, and \(m\) is some new mass parameter in the effective theory. In \cite{Sever:2017ylk} it was shown that the corrections found, including those for the scattering amplitude in the large \(s\), \(t\) limit, correspond to massive endpoint particles on open strings. On the other hand, since the result also applies to theories with only closed strings, such as large \(N\) YM, it was conjectured that the \(m^{3/2}\) correction for closed strings can be associated with the folding points in rotating closed string solutions. There it was argued that the mass terms may arise from higher derivative terms on the worldsheet.
 
The main idea of this work is to add a massive particle at the location of a fold, both for the rotating closed string and the open string with charged endpoints, or any other folded string configuration. With the massive particles the curvature at the folds becomes finite and the eigenmodes are normalizable. We use the mass of the particle as a regulator which we can take to zero at the end of the process in a well defined manner. In \cite{Sonnenschein:2018aqf} we have quantized the open string with massive endpoints for $\frac{m}{TL} \ll 1$ where $m$ is the mass of the endpoint particle, $T$ the tension, and $L$ the length of the open string. The corresponding PS term diverges in the massless limit. We introduced a regularization and renormalization method which is similar to the one used for the ordinary Casimir effect and derived the corresponding intercept. In the limit of zero mass it is equal to the result derived in \cite{Hellerman:2013kba}. It should be emphasized that while the most visible divergence is that of the PS term in the non-critical effective string theory, the problem of normalizing the fluctuation modes around the rotating solution is there whenever we expand around a folded rotating solution, including strings in the critical dimension.

The masses are added by coupling to the string action the worldline action for massive point particles
\be S_{pp} = -\sum_i m_i \int d\tau \sqrt{-\dot X^2}|_{\sigma=\sigma_i} \label{eq:Spp} \ee
at various points \(\sigma_i\) on the string. For the ordinary open string we place these terms at the endpoints of the string, which is well motivated by both theory and phenomenology. In previous works we have explored endpoint masses from the point of view of strings stretched between flavor branes in holographic backgrounds \cite{PandoZayas:2003yb}, and showed that corrections of the form \ref{eq:trajgeneral} are well supported by looking at the Regge trajectories of experimentally observed hadrons \cite{Sonnenschein:2018fph}. The matching with experiment actually goes beyond just the leading term of eq. \ref{eq:trajgeneral}, it also works for large masses where the deviation from linear trajectories is great.

In this paper we argue the role of masses as regulators to the divergences in the semiclassical quantization of fluctuations around rotating string solutions. We argue that to resolve the divergences on folded strings we should place such terms on the folding points of the string, rather than just the boundaries. The reason it works is that the divergences are associated with the fact that the folding points move at the speed of light, and the masses act to slow them down. We add the extra terms by hand, adding the action \ref{eq:Spp} at the points where regularization is required. This might be an unusual way to renormalize, by adding localized counterterms, but the inclusion of mass terms is physically well motivated and in this paper we show it is mathematically consistent.


If we use the massive particle at every folding point as a regulator and take it to zero, we show that the intercept, including the contribution from the PS term, for rotating open strings takes the form
\be a_{\mathrm{open}} = \frac{D-2}{24} + \frac{26-D}{24} = 1 \label{eq:trajopen} \ee
for any \(D\), while for closed strings
\be a_{\mathrm{closed}} = \frac{D-2}{12} + \frac{26-D}{12} = 2 \label{eq:trajclosed}\ee
For both open and closed strings, the result in the massless limit is that the intercept is independent of the dimension, a result which generalizes what was found in \cite{Hellerman:2013kba} to the closed string. This means that for open rotating strings the leading Regge trajectory is
\be J = \alp M^2 + 1 \ee
for any \(D\), and for closed strings
\be J = \frac12\alp M^2 + 2  \ee
also independently of dimension. Here \(\alp = (2\pi T)^{-1}\) for both types of strings, where \(T\) is the string tension. In this paper we also continue the work begun in \cite{Sonnenschein:2019bca}, and analyze the rotating open string with opposite endpoint charges rotating in a magnetic field. As long as the two charges are \(+q\) and \(-q\), such that the total charge of the string is zero, the Regge trajectory is still \(J = \alp M^2 + 1\), independently of either the dimension or the value of the magnetic field.

In this paper we compute the spectrum of fluctuations and the intercept for rotating folded strings with finite masses on their folding points, correcting eqs. \ref{eq:trajopen} and \ref{eq:trajclosed}. The leading correction will be of the form of \ref{eq:trajgeneral}, which we get from the classical solution in the presence of masses. The intercept will include more corrections starting at order \(\left(\frac{m}{TL}\right)^{1/2}\), and these will depend on the mass, as well as the spacetime dimension \(D\), and the external magnetic field for the case of the rotating open string with endpoint charges.

The quantization of the folded rotating closed string and of folded open strings with electric charges on their ends is important for describing hadrons in terms of strings. Glueballs with non-trivial angular momentum are expected to be described by folded closed strings \cite{Sonnenschein:2015zaa} and mesons and baryons are supposed to relate to open string with massive particles and electric charges on their endpoints.

En route toward the renormalization of the folded strings and determination of the intercept we see multiple computations that we can carry out using both the Zeta function renormalization as well as the technique of the Casimir energy contour integral \cite{Lambiase:1995st}. The results match in all the cases discussed. By this comparison we arrive at various identities between infinite sums involving Zeta functions and closed integral expressions which give the same function. These are listed in appendix \ref{app:Zeta}.

This work is organized as follows. In section \ref{sec:folds} we write down the definition and some general properties of folded strings. In section \ref{sec:classical} we present several classical folded solutions of the string equations of motion. These are mainly rotating solutions, for both closed and open strings, and including both flat spacetimes and a few examples of strings rotating in curved backgrounds. We also include an example of a folded non-rotating solution to compare to the rotating solutions.

In section \ref{sec:fluctuations} we discuss in general the semiclassical description of strings, in particular we analyze the fluctuations on rotating strings. Next we perform the semiclassical quantization of the folded closed string (section \ref{sec:quantc}) and the folded open string in a magnetic field (section \ref{sec:quanto}). We describe the quantization in the presence of massive particles on the folding points, and compute the spectrum and the intercept of the theory, in the presence of the folding point masses.

The computation of the intercept also needs a contribution from the Polchinski-Strominger term for non-critical strings. This is addressed is section \ref{sec:PS}. A summary and open questions are in section \ref{sec:summary}.

\section{Folded strings: generalities} \label{sec:folds}
The basic definition and properties of a fold of a bosonic string configuration were written down in \cite{Ganor:1994rm}. In this section we review the basics of folds and further elaborate on folds on rotating strings.

A string configuration $X^\mu(\sigma^a)$ is a map from a world-sheet $\Sigma_g$ of genus $g$  described by a pair of coordinates \((\sigma^0,\sigma^1)\), with $ -\ell\leq\sigma^1\leq \ell$ to  a $d$ dimensional  target spacetime manifold ${\cal M}$ with coordinates $X^\mu$, where $\mu = 0, ... d-1$.

The framework in which we study these maps is the Nambu-Goto (NG) action for which the partition function for closed strings is given by
\be\label{partition function}
{\cal Z}_{\cal M}=\sum_g \lambda_{st}^{2g-2}\int{\cal D} X e^{T\int d^2\sigma \sqrt{\det(h_{ab})}}
\ee
where $\lambda_{st}$ is the string coupling, $T$ is the string tension and  the induced metric $h_{ab}$ is given by
\be
h_{ab}= G_{\mu\nu}(X) \pa_a X^\mu \pa_b X^\nu  
\ee
given a target spacetime with metric $G_{\mu\nu}(X)$. For now we work in Euclidean signature as in \cite{Ganor:1994rm}.

For the case where the target space is two dimensional the NG action can be written as 
\be
S_{NG} = T\int d^2 \sigma \sqrt{det( G_{\mu\nu})}|\det\left (J^\mu_a \right )|
\ee
where $J^\mu_a$ is the Jacobian matrix defined by
\be
J^\mu_a \equiv \frac{\pa X^\mu }{\pa \sigma^a}
\ee
and for the special case of flat target spacetime  
\be
S_{NG} = T\int d^2 \sigma  |\det\left (J^\mu_a \right ) |
\ee

The folded string solutions that we will study in this paper reside on target spaces which are not necessarily two dimensional, however the maps themselves can be described as maps into a two dimensional subspace of the full target spacetime.

The maps $X^\mu(\sigma^a)$ to a two dimensional target space can now be classified in the following way 
\begin{itemize}
\item
Unfolded maps for which
 \be |\det\left(J^\mu_a \right)|\neq 0\ee
 everywhere.
\item
Folds which are curves on the target space for which 
\be\label{2dfold}
|\det\left (J^\mu_a \right )|= 0\ee
at some point (or points) \(\sigma^1=\sigma_f\) in the bulk of the worldsheet $-\ell<\sigma_f<\ell$.
\end{itemize}

In addition to the criterion above that is restricted to two dimensional target space, there are other signatures of folded maps. For a two dimensional target space it is easy to realize that the condition \ref{2dfold} implies also that the determinant of the induced metric vanishes 
\be\label{foldind}
\det h_{ab} =0 
\ee
The induced metric is a $2\times 2$ matrix for any target space dimension so for the more general case one can define a fold using the condition  (\ref{foldind}).
To rewrite the condition in terms of a diffeomorphism invariant quantity we can perform  world sheet coordinates transformations so that the world sheet metric is brought to the following form 
\be\label{wsmetric}
ds^2 = 2h_{+-} d\sigma^+ d\sigma^- 
\ee
The world sheet scalar curvature that corresponds to this metric has the form
\be
{\cal R}_2 = \frac{ \pa_-\pa_+ \ln(h_{+-})}{h_{+-}}
\ee
Thus, on a fold that obeys (\ref{foldind}) we expect the scalar curvature to diverge,
\be |{\cal R}_2|\rightarrow \infty\ee 
at the folding points. The determinant \(\det h_{ab}\) might vanish due to a coordinate singularity, so in general one must verify the singularity associated with the folding point by computing the scalar curvature directly. We will do that for explicit folded solutions in the following section.

For rotating folded strings that will be the focus of this paper there is another physical property that characterizes the fold, and that is that the folding point moves at the speed of light.

Let us assume now the static gauge \(\sigma^0 = \tau = X^0\), and use the remaining repametrization of \(\sigma^1 = \sigma\) to fix \(h_{01} = 0\), such that the induced metric in this gauge is given by
\be h_{ab} = \bmat -1+\beta^2(\tau,\sigma) & 0 \\
0 & X^{\prime2}(\tau,\sigma) \emat \ee
Note that for this analysis we have gone to Lorentzian signature, which we will use for the rest of the paper. Now
\be \beta^2 = \frac{\pa X^i}{\pa \tau} \frac{\pa X^i}{\pa \tau} \ee
is the velocity in target space of any given point. The determinant of the metric is
\be
\det h_{ab} = - (1-\beta^2)(X^\prime)^2
\ee
so it vanishes at points moving at the speed of light, \(\beta^2\to 1\) or at points satisfying \((X^\prime)^2 = \frac{\pa X^i}{\pa\sigma}\frac{\pa X^i}{\pa\sigma} = 0\).

%

We defined above a fold at $-\ell <\sigma_f < \ell$. There are maps where the condition \ref{2dfold} is obeyed at the boundary of the worldsheet coordinates, $\sigma=\ell$ or \(-\ell\). A particular case that will be described in the next section is that of rotating open strings. Then the two conditions
\be \det(h_{ab})=0|_{\sigma=\pm \ell}\qquad \beta^2=1|_{\sigma=\pm\ell} \ee
are met, but the maps are not folded, since this happens at the boundary.

In the next section we write various particular solutions of the string equations of motion and analyze the induce metric, velocity, and worldsheet curvature on those solutions to demonstrate the ideas in this section.

\section{Classical folded solutions of the string equations of motion} \label{sec:classical} 
In certain circumstances the  classical string equations of motion admit folded string solutions. We focus mainly on rotating solutions. These include rotations in flat spacetime of closed strings and open strings in an external magnetic field, which are solutions that must include folds. We briefly describe rotating folded strings in $AdS$ target space, as well as closed folded strings in generic holographic confining backgrounds. In the last part of the section we describe an example of a non-rotating folded solution.

\subsection{Rotating folded string solutions in flat spacetime}
The action of the bosonic string is the Nambu-Goto action given by
\be S = -\frac{1}{2\pi\alp}\int d\tau d\sigma \sqrt{-\det h_{ab}} \,,\ee
where
\be h_{ab} = G_{\mu\nu}(X) \pa_a X^\mu \pa_b X^\nu \ee
is the induced metric on the world sheet and
\be \alp = \frac{1}{2\pi T} \,. \ee
where $T$ is the tension of the string. We assume for now that the background is flat \(G_{\mu\nu}(X) = \eta_{\mu\nu}\).

The equations of motion derived from the NG action read
\be \pa_\alpha(\sqrt{-h} h^{\alpha\beta}\pa_\beta X^\mu)=0
\ee

The rotating configuration given by
\be X^0 = \tau \qquad X^1 = R(\sigma)\cos(\omega\tau) \qquad X^2 = R(\sigma)\sin(\omega\tau) \label{eq:rotsolgen}\ee
is always a solution of the equations of motion. We pick the form
\be R(\sigma) = \frac{1}{\omega}\cos(\omega\sigma+\phi) \ee
such that the solution also obeys the orthogonal gauge conditions (or Virasoro constraints):
\be \dot X^2 + X^{\prime2} = \dot X\cdot X^\prime = 0 \ee
Or in other words the worldsheet metric is conformally flat, \(h_{ab} = e^\varphi \eta_{ab}\).

Imposing different boundary conditions on the string will fix the allowed values of the parameters \(\omega\) and \(\phi\) that appear in the solution.

The induced metric on the rotating solution is
\be h_{ab} = \bmat -1+\omega^2 R^2 & 0 \\ 0 & R^{\prime2} \emat \ee
When \(R(\sigma) = \frac1\omega\cos(\omega \sigma+\phi)\), it is
\be h_{ab} = \sin^2(\omega\sigma+\phi) \eta_{ab} \ee
The velocity of a given point on the string is given by
\be \beta^2(\sigma) = \omega^2 R^2 = \cos^2(\omega\sigma+\phi) \qquad \gamma(\sigma) = \frac{1}{\sqrt{1-\beta^2}} = \frac{1}{|\sin(\omega\sigma+\phi)|} \ee
and the worldsheet curvature is
\be \mathcal R_2(\sigma) = \frac{2\omega^2}{\sin^4(\omega\sigma+\phi)} = 2\omega^2\gamma^4(\sigma) \ee
The curvature diverges at points that move at the speed of light, \(\beta^2 = 1\).

\subsubsection{Rotating closed strings in flat spacetime} \label{sec:classicalFoldedString}
In the case of the closed string, the worldsheet coordinates are \(\tau \in (-\infty,\infty)\) and \(\sigma \in (-\ell,\ell)\)
The boundary conditions for a closed string are the periodic identification \(\sigma \sim \sigma+2\ell\), in particular \(-\ell\sim \ell\). The rotating string solution is 
\be X^0 = \tau \qquad X^1 = \frac{1}{\omega}\cos(\omega\sigma)\cos(\omega\tau) \qquad X^2 = \frac{1}{\omega}\cos(\omega\sigma)\sin(\omega\tau) \,.\label{eq:rotsolc}\ee
where \(\omega\) takes the values
\be \omega = n\pi/\ell \ee
for any integer \(n\).

The energy of this configuration is
\be E = T \int_{-\ell}^\ell d\sigma \partial_\tau X^0 = 2T\ell \ee
The angular momentum, going to polar coordinates in the 12 plane,
\be J = T \int_{-\ell}^\ell d\sigma \rho^2 \partial_\tau \theta =
\frac{T}{\omega} \int_{-\ell}^\ell d\sigma \sin^2(\omega\sigma) = \frac{n T\ell^2}{\pi}\ee
From the last two equations we can easily see that for the classical rotating folded string
\be J = \frac{1}{4\pi T n}E^2 = \frac{1}{2n}\alp E^2 \ee
The slope of the trajectory is \(\frac1{2n}\alp_{o}\), where \(\alp_o = \frac{1}{2\pi T}\) is the Regge slope of an ordinary open string trajectory of the same string tension. We can say that the tension of the string is effectively $2n T$, which is a property derived from the fact that the rotating closed string is a folded string. In particular, \(n=1\) gives the \emph{leading} trajectory, maximizing \(J\) for a given energy \(E\), and based on that we say that the closed string slope is double that of the open string, \(\alp_c = \frac12\alp_o\).

\begin{figure}[ht!] \centering
\includegraphics[width=0.50\textwidth]{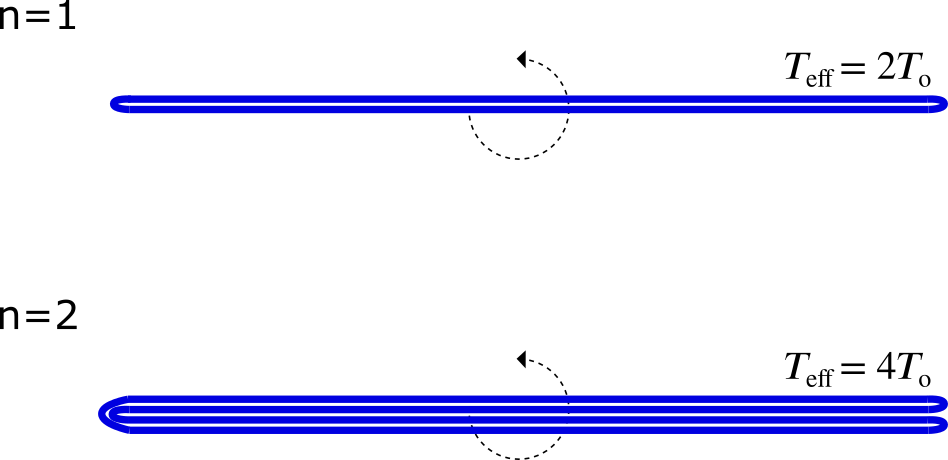}
\caption{\label{fig:closed_folded} The rotating folded string has \(2n\) folding points and effective tension \(2n T_o\), where \(T_o\) is the bare tension, or the tension of a non-folded open string. The solutions are drawn with finite width, but the classical solutions we write have zero width, such that the different segments and the folding points coincide.}
\end{figure}

For the general \(n\)-folded solution
\be \beta^2(\sigma) = \cos^2(\frac{n\pi}{\ell}\sigma) \qquad\qquad \mathcal R_2 = \frac{2\omega^2}{\sin^4(\frac{n\pi}{\ell}\sigma)} \ee
For \(n=1\) there are two folding points at \(\sigma = 0\) and \(\pm\ell\) which move at the speed of light, and the worldsheet curvature is divergent on them. In general there are \(2n\) points, \(\sigma_k = \frac{\ell}{n}(k-n)\) with \(k=0,\ldots,2n-1\) which are folding points along the string. All the \(\sigma_k\) with even \(k\) coincide in target space with \(\sigma=0\), while all the odd \(k\) coincide with \(\sigma = \pm\ell\), so in effect we have \(n\) copies of the original folded solution sitting on top of each other. See figure \ref{fig:closed_folded}. From a target space point of view, one can say there are two folding points, and the string passes through each of them \(n\) times. Classically this distinction is unimportant but quantum fluctuations around the multiply folded solution will need to have defined ``boundary'' conditions at each of the \(2n\) folding points.

Most of this paper will be focused on the leading solution with \(n=1\). The ordinary quantization of the closed bosonic string in critical dimensions leads to the leading Regge trajectory \(J = \frac12\alp + a\), with the intercept \(a = 2\), and in we extend this result to non-critical dimensions by the semiclassical quantization around the folded rotating solution.

%
%
 
\subsubsection{Rotating closed strings in two planes of rotation} \label{sec:classicalTwoPlanes}
For \(D \geq 5\), the rotation group \(SO(D-1)\) contains \(SO(4)\sim SU(2)\times SU(2)\), so the solution written in the previous subsection is not the most general rotating solution, since we can write a more general one with two independent angular momentum numbers. A generic rotating string solution with angular momenta in two different planes can be written as
\[ X^0 = \tau \qquad X^1 = \frac{\cos\xi}{\omega}\sin(\omega\sigma)\cos(\omega\tau) \qquad X^2 = \frac{\cos\xi}{\omega}\sin(\omega\sigma)\sin(\omega\tau) \qquad  \]
\be X^3 = \frac{\sin\xi}{\tilde\omega}\sin(\tilde\omega\sigma-\phi)\cos(\tilde\omega\tau) \qquad X^4 = \frac{\sin\xi}{\tilde\omega}\sin(\tilde\omega\sigma-\phi)\sin(\tilde\omega\tau) \,.\label{eq:rotsol2}\ee
The equations of motion and Virasoro constraints are obeyed for any choice of the parameters.

For the closed string \(\omega\) and \(\tilde\omega\) are integer multiples of \(\pi/\ell\).  As before, we are interested in the leading solution which maximizes angular momentum for a given energy. Therefore we take \(\omega = \tilde\omega = \frac{\pi}{\ell}\). This solution does not have folds. In addition, it does not develop folds when looking at solutions where the angular velocities are given by higher multiples of \(\frac\pi\ell\).

The mass and angular momentum are
\be M = \frac{\pi}{2} TL \qquad J_1 \equiv J_{12} = \frac{\pi}{16}TL^2\cos^2\xi  \qquad J_2 \equiv J_{34} = \frac{\pi}{16}TL^2\sin^2\xi  \ee
where \(L\) is the length of the string, and the closed string Regge trajectory is now given by
\be J_1 + J_2 = \frac{1}{4\pi T}M^2 = \frac12 \alp M^2 \ee
plus an intercept from quantum corrections.

The parameters \(\xi\) and \(\phi\) are not fully independent for \(\omega=\tilde\omega\), since we can change them by performing rotations in the \(13\) and \(24\) planes. If the rotation is done simultaneously and with the same angle in both planes we get another solution of the same form as in eq. \ref{eq:rotsol2} (up to a shift in \(\sigma\)). If we start from any solution with \(\phi\neq0\) and \(\xi\neq0\), we can rotate by an appropriate angle to a solution with \(\phi=\pi/2\). On the other hand, a solution with \(\phi=0\) can be rotated to give back the solution with \(\xi=0\), which is the folded solution rotating in a single plane that we examined above.

Therefore, we now look at the solution with \(\phi=\frac\pi2\) and \(\omega = \tilde\omega = \pi/\ell\):
\[ X^0 = \tau \qquad X^1 = \frac{\cos\xi}{\omega}\sin(\omega\sigma)\cos(\omega\tau) \qquad X^2 = \frac{\cos\xi}{\omega}\sin(\omega\sigma)\sin(\omega\tau) \qquad  \]
\be X^3 = \frac{\sin\xi}{\omega}\cos(\omega\sigma)\cos(\omega\tau) \qquad X^4 = \frac{\sin\xi}{\omega}\cos(\omega\sigma)\sin(\omega\tau) \,.\label{eq:rotsol2b}\ee
We can show now that as long as \(J_1\) and \(J_2\) are both non-zero, then there is no fold, no point moving at the speed of light, and everything is finite.

The induced metric on the worldsheet of this solution is
\be h_{ab} = \frac12[1-\cos(2\xi)\cos(2\omega\sigma)]\eta_{ab} \ee
We can define \(J_\pm = J_1 \pm J_2\), in terms of which
\be h_{ab} = \frac{1}{2J_+}[J_+-J_-\cos(2\omega\sigma)]\eta_{ab} \ee
The velocity of a given point along the string is
\be \beta^2(\sigma) = \frac{1}{2J_+}(J_+ + J_- \cos(2\omega\sigma)) \ee
The worldsheet curvature,
\be \mathcal R_2 = 8\omega^2 J_+ J_- \frac{J_- - J_+\cos(2\omega\sigma)}{(J_+ - J_-\cos(2\omega\sigma))^3} \ee
is finite everywhere as long as \(|J_+| \neq |J_-|\), which is the case as long as both \(J_1\) and \(J_2\) are non-zero. If one of them is zero, we are back to the case of the folded string, with the folding points \(\sigma = 0\) and \(\ell\) moving at the speed of light and having divergent worldsheet curvature.

If one of the angular momenta is much larger than the other, w.l.o.g. \(J_1 \gg J_2\), we can get arbitrarily close to having a folding point by increasing the ratio \(J_1/J_2\). The velocity of the extremal points in that case will be
\be \beta(0) = \beta(\ell) = 1- \frac{J_2}{J_1} + \ldots \ee
while the worldsheet curvature at these points is
\be \mathcal R_2 = -2\omega^2 \left(\frac{J_1}{J_2}\right)^2 +\ldots \label{eq:curvTwoPlanes}\ee
We see that for a finite ratio of \(J_1/J_2\) the curvature is large and negative at the extremal points. On the other hand, the curvature at the folding points in the \(J_2 = 0\) limit is positive infinity.

We will not quantize the fluctuations around this solution in this paper, but we discuss some aspects of its quantization, which was carried out in \cite{Hellerman:2013kba,Zahn:2016hxw}, in section \ref{sec:PSc2}.


\subsection {Rotating open string solutions}
\begin{figure}[ht!] \centering
\includegraphics[width=0.60\textwidth]{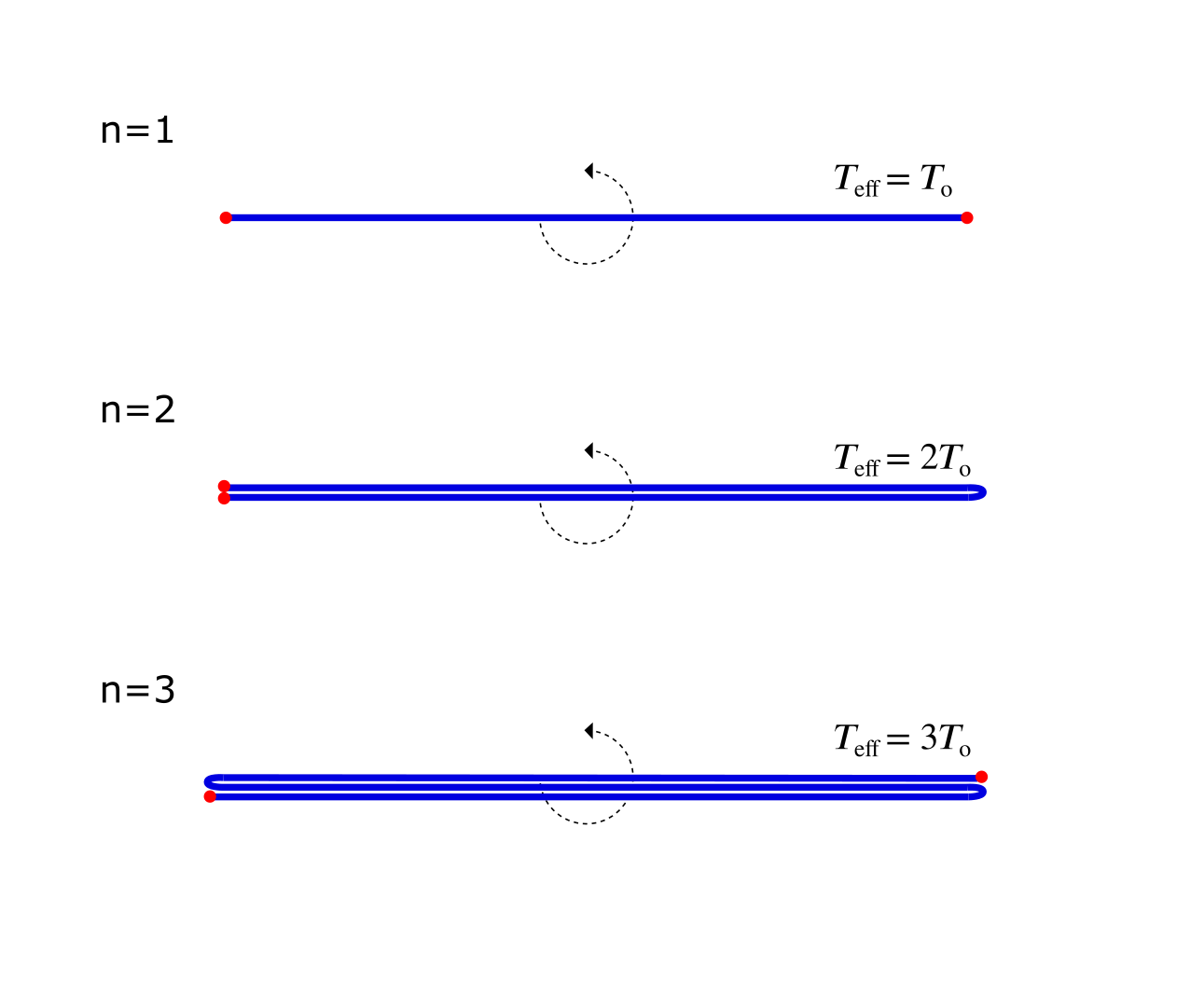}
\caption{\label{fig:open_folded} The ordinary rotating string has no folds, but for \(\omega=n\frac\pi\ell\) with \(n>1\) there will be \(n-1\) folding points and effective tension \(n T_o\). For even \(n\) the endpoints of the string coincide in target space.}
\end{figure}

For the open string we take \(\sigma\in(0,\ell)\). For open strings with Neumann boundary conditions, there exist rotating solutions of the same form as the closed string solution of eq. \label{eq:rotsol} and with \(\omega=\frac{n\pi}{\ell}\). These solutions have the Regge trajectories
\be J = \frac{1}{2n\pi T}M^2 = \frac{1}{n}\alp M^2 \ee
The leading trajectory with \(n = 1\) is not folded. The endpoints \(\sigma = 0\) and \(\ell\) move at the speed of light, and the problem of the semiclassical quantization of this system in non-critical dimensions, with appropriate treatment of divergences at the boundaries, was addressed in \cite{Hellerman:2013kba,Sonnenschein:2018aqf}. For general \(n>1\), there are \(n-1\) folding points, \(\sigma_k = \frac{\ell}{n}k\) with \(k=1,\ldots,n-1\). For even \(n\) the string folds back on itself such that the endpoints \(0\) and \(\ell\) coincide in target space and there is a folding point at the other end of the rotating string. For odd \(n\) they sit at the opposite ends. See figure \ref{fig:open_folded}.

\subsubsection{Rotating open strings in magnetic fields} \label{sec:rot_open_mag}
\begin{figure}[ht!]
\begin{center}
\includegraphics[width=0.48\textwidth]{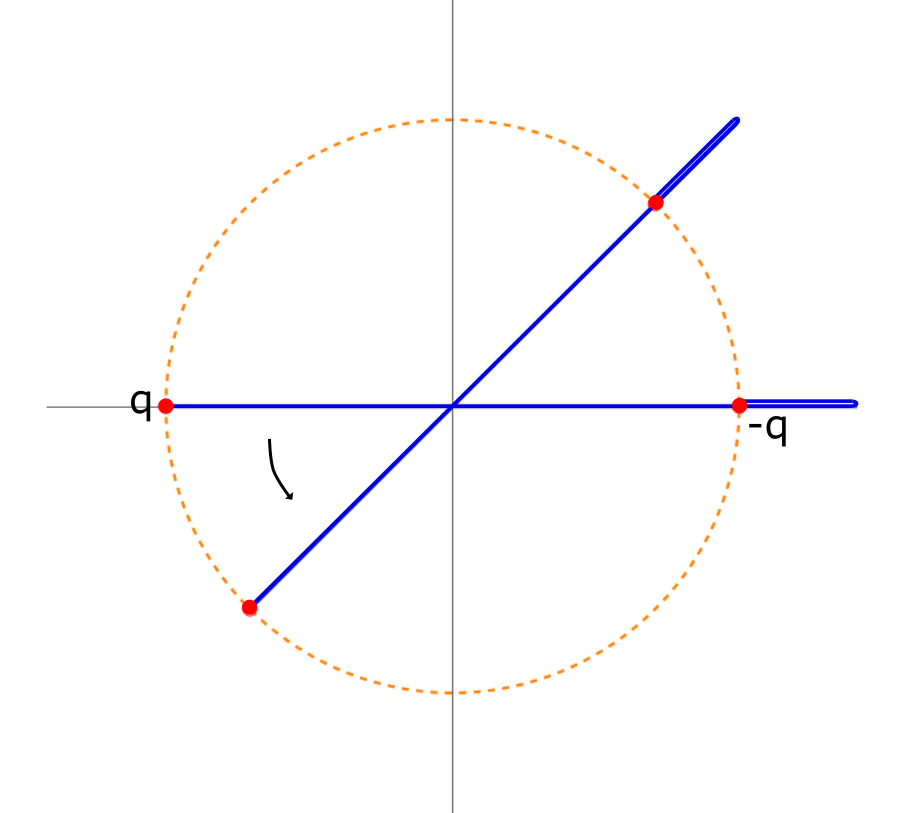}
  \caption{\label{fig:rotating_neutral} The rotating solution of a neutral string with endpoint charges in a magnetic field. The charges rotate around the central point between them. The magnetic field is in the \(z\) direction, coming out of the plane, and the rotation is counterclockwise.}
 \end{center}
\end{figure}

Instead of discussing the folded \(n>1\) solutions of the Neumann open string, we turn to another system where folds must develop even for the leading solution with \(n=1\). This is the open string coupled to an electromagnetic field whose action is the Nambu-Goto action plus the boundary terms
\be S_q = q \int d\tau A_\mu \dot X^\mu|_{\sigma=0} - q\int d\tau A_{\mu} \dot X^\mu|_{\sigma=\ell} \ee
such that the boundary conditions of the string are
\be T X^{\prime \mu} + q F^\mu{}_\nu \dot X^\nu = 0 \ee
When we take a purely magnetic field, that is \(F_{12} = -F_{12} = B\) we find that the rotating solution now has a fold even for \(n=1\). Note that while here we discuss the specific case that the charges on the endpoints are \(+q\) and \(-q\), more generally it holds that if there is positive charge on one endpoint and a negative charge on the other, then any rotating solution must have a fold, as in figure \ref{fig:rotating_neutral}. This is because the Lorentz force acting on the endpoint charges will be in the same direction for both charges, and the string tension that balances it will have to act in the same direction on both endpoints. This is impossible without a fold. See figure \ref{fig:rotating_neutral_fold}.

We assume w.l.o.g. that \(B>0\). The solution we discuss is again of the form
\be X^0 = \tau \qquad X^1 = \frac{1}{\omega}\cos(\omega\sigma+\phi)\cos(\omega\tau) \qquad X^2 = \frac{1}{\omega}\cos(\omega\sigma+\phi)\sin(\omega\tau) \ee
where the boundary conditions dictate that
\be \omega = \frac\pi\ell \qquad \phi = \arctan(\frac{q B}{T}) \ee
We took the leading solution (\(n=1\)) with no additional folding points.

For this solution the charged endpoints of the string move at a finite velocity given by \(\beta_q^2 = \cos^2\phi\), while the folding point located at \(\sigf = \ell(1-\frac{\phi}{\pi})\) moves at the speed of light. Because of this, the worldsheet curvature is no longer divergent at the endpoints, the boundary of the worldsheet, but only at the folding point which is now in the bulk.

The coupling to the magnetic field does not change the Regge trajectory, as the energy and angular momentum are the same as for the Neumann string
\be E = T \ell \qquad  J = \frac{T\ell^2}{2\pi} \ee
with the Regge trajectory
\be J = \alp E^2 \ee
In the critical dimension, the system can be quantized exactly, with the result \(J = \alp M^2 + 1\) \cite{Abouelsaood:1986gd}. We will compute the intercept in the semiclassical quantization around the rotating string with the fold, and show that \(a=1\) is the result for any dimension \(D\), which is the same as the result for the rotating open string with no external field.
     
\subsubsection{Rotating open string with massive endpoints} \label{sec:classicalMass}
We will make use in later sections of a rotating sting with massive endpoints. This solution is not folded and does not exhibit any divergences. It will be the basis for the regularization of the divergences associated with the folds. The action of the system of an open string with massive endpoints is given by adding to the string action
\be S = S_{st} + S_{pp}\vert_{\sigma=0} + S_{pp}\vert_{\sigma=\ell} \ee
the mass terms
\be S_{pp} = -m\int d\tau \sqrt{-\dot X^2} \ee
For now we assume equal endpoint masses. The bulk equations of motion are the same as those of the ordinary open string but the boundary conditions now are
\be T\sqrt{-h}\pa^\sigma X^\mu \pm m \pa_\tau\left(\frac{\dot X^\mu}{\sqrt{-\dot X^2}}\right) = 0 \qquad \sigma = 0,\ell\ee
with the plus sign at \(0\) and minus at \(\ell\). For the rotating solution, the requirement now is that
\be T\frac{\sqrt{(1-\omega^2R^2)R^\prime{}^2}}{R^\prime} \mp m \frac{\omega^2R}{\sqrt{1-\omega^2R^2}} = 0 \qquad \sigma = 0,\ell \ee
When expressed in terms of the endpoint velocity \(\beta\) and length of the string \(L\) the boundary conditions read 
\be \frac{T}{\gamma} = \frac{2\gamma m \beta^2}{L} \qquad \Rightarrow \qquad \frac{TL}{2m} = \gamma^2\beta^2 \label{eq:bdc}\ee
This equation has the simple interpretation of the balancing between the tension and centrifugal  force acting on the endpoint particle.
 
The classical energy and angular momentum are expressible as functions of \(T\), \(m\), \(L\) and \(\beta\):
\begin{align} E &= \frac{2m}{\sqrt{1-\beta^2}} + TL\frac{\arcsin\beta}{\beta} \label{eq:classical_E}\\
J &= \frac{m L \beta}{\sqrt{1-\beta^2}}+\frac14TL^2\frac{\arcsin\beta-\beta\sqrt{1-\beta^2}}{\beta^2} \label{eq:classical_J}\end{align}
And we find the mass corrected Regge trajectories
\be J = \alp E^2\left(1 - \frac{8\sqrt\pi}{3}\left(\frac m E\right)^{3/2} + \frac{2\pi^{3/2}}{5}\left(\frac m E\right)^{5/2} +\ldots \right) \ee
 It is easy to see that the massive endpoints now rotate at a finite velocity, which can be expressed as
\be
\beta^2 = \left (1+ \frac{2m}{TL} \right )^ {-1} = \frac{TL}{2m+TL}
\ee
The worldsheet curvature has a finite maximum now at the endpoints which can be written as
\be \mathcal R_2 = 2\gamma^4 \omega^2 = \frac{8 \gamma^4 \beta^2}{L^2} = \frac{8}{L^2}\frac{TL}{2m}\left(1+\frac{TL}{2m}\right)\ee
At small masses or high energies, i.e. when \(m \ll E\), we can write
\be \mathcal R_2 = \frac{2 T^2}{m^2}\left(1 + \frac{\pi m}{E} + \ldots\right) \ee
so the curvature remains finite even as we take the energy of the string to infinity, as long as we started with a finite mass.


\subsubsection{Rotating strings in magnetic field with massive endpoints}
As discussed in \cite{Sonnenschein:2019bca}, the inclusion of mass terms in addition EM charges can prevent the creation of the fold in the rotating solution.

For the ansatz of a rotating solution as in eq. \ref{eq:rotsol}, the boundary conditions, where we have massive endpoints of mass \(m\) with charges \(+q\) at \(\sigma=0\) and \(-q\) at \(\sigma=\ell\), are
\begin{align} &-T \sin\phi + \frac{m\omega \cos\phi}{|\sin\phi|} + q B \cos\phi = 0 &\qquad \sigma = 0 \nonumber \\  &T \sin(\omega\ell+\phi) + \frac{m\omega \cos(\omega\ell+\phi)}{|\sin(\omega\ell+\phi)|} - q B \cos(\omega\ell+\phi) = 0 &\qquad \sigma = \ell \label{eq:bdm} \end{align}
When \(m = 0\) we find the folded solution with \(\omega\ell=n\pi\) and \(\phi = \arctan(q B/T)\). In general we can assume that \(qB>0\) and \(\omega>0\). In that case for finite masses we can find a solution with \(0\leq\phi\leq\frac\pi2\).

\begin{figure}[h]
\begin{center}
\includegraphics[width=0.48\textwidth]{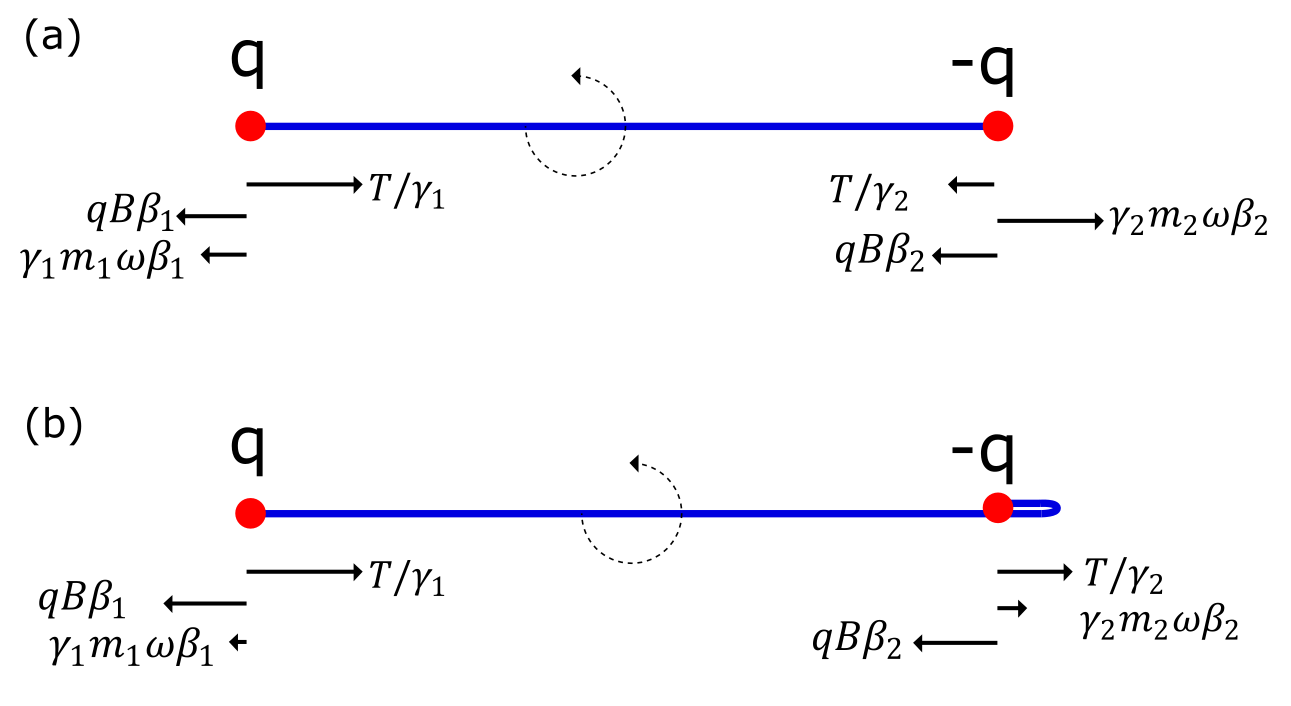}
  \caption{\label{fig:rotating_neutral_fold} The forces acting on the endpoint particles when the fold develops. The forces are drawn to scale for a solution with \(qB/T=1/8\) and equal endpoint masses. In (a) we take \(m/T\ell=0.003\) (\(T\ell\) being equal to the energy carried by the string) and we find a solution without a fold. When the masses are taken to be even smaller, \(m/T\ell\) = 0.001 then we have only the solution with the fold.}
 \end{center}
\end{figure}

With masses, there are two possible types of solutions. The Lorentz force on the charges is in the same direction for both points. In the absence of masses, the fold is necessary since we need the tension to balance the Lorentz force. With masses, the centrifugal force can balance the two other forces with no fold required. This is depicted in figure \ref{fig:rotating_neutral_fold}.

We can write everything in terms of dimensionless quantities. Define \(\eta = \frac{m}{T\ell}\), \(b =\frac{qB}{T}\), and \(\delta=\omega\ell\), so
\begin{align} &-\sin\phi + \eta\frac{\delta\cos\phi}{|\sin\phi|} + b \cos\phi = 0 &\qquad \sigma = 0 \nonumber \\  & \sin(\delta+\phi) + \eta\frac{\delta \cos(\delta+\phi)}{|\sin(\delta+\phi)|} - b \cos(\delta+\phi) = 0 &\qquad \sigma = \ell \label{eq:bdm2} \end{align}
The solution will have a folding point if \(\delta+\phi \geq \pi\). 

For zero masses and any finite \(b\) the solution must have a fold as stated in section \ref{sec:rot_open_mag}, but when we start to increase the mass at the endpoints, we reach a point where a solution without a fold exists. The two possibilities are depicted in figure \ref{fig:rotating_neutral_fold}.

\subsection{ Rotating folded closed strings in holography}
Immediately  after the pioneering papers of the AdS/CFT  correspondence solutions of classical equations of motion of folded closed strings in both the $AdS_5$ part of the target space as well as the $S^5$ part have been written \cite{Gubser:2002tv} and analyzed later using a semiclassical quantization procedure \cite{Beccaria:2010ry}. The solutions were then generalized to other holographic backgrounds and in particular to \emph{confining} holographic backgrounds \cite{PandoZayas:2003yb}. In the next subsection we write the rotating solutions for a closed string in $AdS$ and then in the next subsection we present a solution for a generic holographic background.

\subsubsection{Rotating folded strings in AdS}
In \cite{Gubser:2002tv} a solution of a folded closed string in $AdS$ was written down. The solution is given in terms of the global coordinates\footnote{Here we follow the analysis of \cite{Beccaria:2010ry}.}
\be ds^2 = -\cosh^2\!\rho\, dt^2 + d\rho^2 + \sinh^2\!\rho\, (d\phi^2 + \sin^2\phi d\Omega_{d-3}) \ee
The solution for a closed string, \(\sigma \in [0,2\pi]\), in a single plane of rotation is
\be
t = \kappa\tau \qquad \theta= \omega\tau \qquad \rho = \rho(\sigma)= \rho(\sigma + 2\pi)
\ee
where $\rho(\sigma)$ has to solve the equation of motion (in conformal gauge)
\be
(\rho^\prime)^2 = \kappa^2\cosh^2\rho-\omega^2\sinh^2\rho
\ee
The solution of the equation of motion can be written as follows
\be
\sinh\rho(\sigma)=\frac{k}{\sqrt{1-k^2}}\mathrm{cn}(\omega\sigma + {\cal K}|k^2) \qquad  \rho^\prime(\sigma) = \kappa \mathrm{sn}(\omega\sigma + {\cal K}|k^2) 
\ee
where cn and sn are the Jacobi elliptic functions, and ${\cal K}\equiv {\cal K}(k^2)$ is the complete elliptic integral of the first kind with elliptic modulus given by $k=\frac{\kappa}{\omega}$. The radial coordinate $\rho$ varies from \(\rho(0) = 0\) to its maximal value  $\rho_0$ which is given by 
\be
\coth^2(\rho_0) = \frac{1}{k^2}
\ee
The periodicity condition implies that  ${\cal K}= \frac{\pi \omega}{2}$.

The induced metric on this solution is given by
\be h_{ab} = \rho^\prime(\sigma)\eta_{ab} \ee
and the scalar curvature is
\be
{\mathcal R}_2 = -2 + \frac{2\kappa^2 \omega^2}{(\rho^\prime(\sigma))^4}
\ee
The difference from the rotating string in flat spacetime being the constant part of \(-2\).

Considering the two dimensional sub manifold of the target space spanned by the coordinates $(t, \rho$), the matrix $J^\mu_\alpha$ reads 
\be
J^\mu_\alpha = 
\bmat \pa_\tau t & \pa_\tau \rho \\ \pa_\sigma t  & \pa_\sigma \rho  \emat = 
\bmat 1 &  0 \\ 0 & \rho^\prime(\sigma)  \emat
\ee
Thus at  the zeros of $\rho^\prime(\sigma) = \mathrm{sn}( \omega\sigma + {\cal K}|k^2)$, which are $\sigma= \frac{\pi}{2}$ and  $\sigma= \frac{3\pi}{2}$ the determinant of  $J^\mu_\alpha$ vanishes. The solution is folded at the two points and the worldsheet curvature diverges there.

\subsubsection{Rotating folded strings in confining backgrounds} \label{sec:classicalConf}
The analysis of the quantum  rotating folded closed  string in confining background  was performed in \cite{PandoZayas:2003yb}. Here we briefly summarize the classical solution. Assume a five dimensional confining background  of the form
\be
ds^2 = G(r)^{-1/2}\eta_{\mu\nu} dx^\mu dx^\nu + G(r)^{1/2} dr^2
\ee
where $r$ is the holographic direction and \(\mu,\nu = 0,1,2,3\). Following \cite{Kinar:1998vq} a sufficient condition for this background to be a confining one, namely that a  rectangular Wilson loop admits area law behavior, is that there exists some \(r_0\) such that
\be \label{confcond}
\pa_r G(r) =0|_{r=r_0} \qquad  \mathrm{and} \qquad   G(r_0)>0
\ee
The equations of motion for a string in this background are 
\be 
\pa_a ( G(r)^{-1/2} \eta^{ab}\pa_b X^\mu) = 0
\ee
\be
\pa_a ( G(r)^{1/2} \eta^{ab}\pa_b r) =  \frac12 \pa_r G(r)^{-1/2}\eta^{ab}\pa_X^\mu\pa_b X^\nu \eta_{\mu\nu}  
\ee

We search now for a rotation solution. As before, we look at a closed string with \(\sigma \in [-\ell,\ell]\). We take for it the following ansatz  
\be X^0 =  \tau \qquad X^1 =\frac{1}{\omega} g(\sigma) \cos(\omega \tau)  \qquad X^2 =\frac{1}{\omega} g(\sigma)  \sin(\omega t) \nonumber \ee
\be X^3 = 0 \qquad r= r(\sigma)  \label{eq:rotsolconf}
\ee
with \(\omega = \frac\pi\ell\). For this configuration the radial equation of motion and the Virasoro constraint read
\be
\pa_\sigma(G(r)^{1/2} r') -\frac12\pa_r(G(r)^{-1/2}) (1-\w^2 g^2 - (g')^2) = 0 ] \label{eq:361} \ee
\be G(r)^{1/2}(r')^2-G(r)^{-1/2}(1-\w^2 g^2 - (g')^2) = 0 \label{eq:362}
\ee
Invoking now the condition of confining background (\ref{confcond}) we find that for $r(\sigma)= r_0$  the first equation is solved and that the rotating solution (\ref{eq:rotsolconf}) solves the second equation as well.
Substituting the solution into the expression of the energy and angular momentum we find 
\be
 J= \frac12  {\alp}G(r_0)^{1/2} E^2
\ee
namely that the tension of the rotating string located at \(r=r_0\) is $T_{\mathrm{eff}}=\frac{G(r_0)^{-1/2}}{2\pi \alp}$.

Since this solution is located at constant \(r_0\), it is almost identical (at least classically) to a rotating string in flat spacetime, with expressions differing by appropriate factors of \(G^{1/2}(r_0)\). The induced metric on the worldsheet is now, picking the parametrization \(g(\sigma) = \cos(\omega\sigma)\),
\be h_{ab} = G^{-1/2}(r_0)\sin^2(\omega\sigma)\eta_{ab} \ee
and the worldsheet curvature
\be \mathcal R_2 = \frac{2G^{1/2}(r_0)\omega^2}{\sin^4(\omega\sigma)} \ee
The velocity in the four dimensional space is defined by
\be 1-\beta^2 = -\eta_{\mu\nu}\frac{dX^\mu}{dX^0}\frac{dX^\nu}{dX^0} \ee
and is still
\be \beta^2 = \cos^2(\omega\sigma) \ee

\subsection{Folded non-rotating string solutions} \label{sec:classicalMaldacena}
To extend the scope of the  treatment of folded string configurations, let us look at another example of a folded solution found in \cite{Maldacena:2005hi} for a linear dilaton system in two dimensional spacetime. A related solution was analyzed recently in \cite{Attali:2018goq}.

In terms of the worldsheet coordinates $\tau$ and $\sigma$ the folded solution reads
\be \label{foldst} 
X^0  = \tau\,, \qquad
X^1  =  x_0- Q \log\left [ \cosh\left ( \frac{\sigma}{Q}\right )+ \cosh\left ( \frac{\tau}{Q}\right )\right]
\ee
Here one considers infinitely long strings, \(\sigma \in(-\infty,\infty)\). In terms of worldsheet light cone coordinates the solution is
\be \label{fold+-}
X^0  = \sigma^+ + \sigma^-\,,\qquad
X^1  =  \tilde x_0- Q \left [\log \cosh\left ( \frac{\sigma^+}{Q}\right )+ \log \cosh\left ( \frac{\sigma^-}{Q}\right ) \right ] \ee

It is clear from (\ref{foldst}) that there is a fold at $\sigma=0$. Using the definition of a fold given in (\ref{2dfold})  we examine the matrix $J^\mu_\alpha$ given by
\be
J^\mu_\alpha=
\begin{bmatrix}
1& -\frac{ \sinh\left(\frac{\tau}{Q} \right)}{\left [ \cosh\left ( \frac{\sigma}{Q}\right )+ \cosh\left ( \frac{\tau}{Q}\right ) \right ]} \\
0& -\frac{ \sinh\left(\frac{\sigma}{Q} \right)}{\left [ \cosh\left ( \frac{\sigma}{Q}\right )+ \cosh\left ( \frac{\tau}{Q}\right ) \right ]}
\end{bmatrix}
\ee
and we see that its determinant vanishes at $\sigma=0$, and there is a fold in the string solution. 

The induced metric associated with this fold solution is given by 
\be
h_{\tau\tau}= 1- \frac{\sinh^2{\frac{\tau}{Q}}}{(\cosh{\frac{\tau}{Q}} + \cosh{\frac{\sigma}{Q}})^2} \qquad h_{\tau\sigma}= -\frac{\sinh{\frac{\sigma}{Q}}\sinh{\frac{\tau}{Q}}}{(\cosh{\frac{\tau}{Q}} + \cosh{\frac{\sigma}{Q}})^2}  \qquad h_{\sigma\sigma}= -\frac{\sinh^2{\frac{\sigma}{Q}}}{(\cosh{\frac{\tau}{Q}} + \cosh{\frac{\sigma}{Q}})^2} 
\ee
or in light-cone worldsheet coordinates 
\be
h_{++} = \frac{1}{\cosh^2{\frac{\sigma^+}{Q}}}\qquad 
h_{--} = \frac{1}{\cosh^2{\frac{\sigma^-}{Q}}}\qquad 
h_{+-} = 1-\tanh{\frac{\sigma^+}{Q}}\tanh{\frac{\sigma^-}{Q}}
\ee
It is easy to realize that at the folding point \(\sigma = 0\), the determinant vanishes $\det(h_{\alpha\beta})=0$. The worldsheet curvature turns out to be \({\cal R}_2 =0\) at all other points. On the fold $\sigma=0$ there is a localized singularity where $\det(h_{\alpha\beta})=0$. We do not find a singularity of \(\mathcal R_2\) there, but upon quantization one should be careful around the folding point.

As for the velocity of the fold, it is
\be
\beta(\tau) = \frac{\pa X^1}{\pa X^0}|_{\sigma=0}= -\frac{ \sinh\left(\frac{\tau}{Q} \right)}{1+ \cosh\left ( \frac{\tau}{Q}\right )}
\ee 
The velocity of the fold is finite, except in the limit in the limit of $\tau\rightarrow \infty$ where it accelerates to speed of light $|\beta|\rightarrow 1$.  We have in this solution an example of a fold that is moving at a finite velocity, and while the worldsheet is flat everywhere else, it is singular at the folding point.

\section{Quantum fluctuations on strings with folds} \label{sec:fluctuations}
In this section we will see how the divergences we encounter in the classical folded string solutions affect quantum fluctuations when attempting the semiclassical quantization of the string. We write the general quadratic action for the fluctuations, then move to the particular case of a rotating string.
\subsection{Expanding around classical solutions}
Starting from the Nambu-Goto action
\be S_{NG} = -T\int d^2\sigma \sqrt{-\det h_{ab}} \ee
we expand around some ``background'' by writing \(X^\mu = \bar X^\mu + Y^\mu\) with \(\bar X^\mu\) some solution to the classical equations of motion. We can denote the classical part of the induced metric as
\be g_{ab} = G_{\mu\nu}(\bar X) \pa_a \bar X^\mu \pa_b \bar X^\nu \ee
where \(G_{\mu\nu}\) is the spacetime background metric. We define the fluctuations of the metric as
\be y_{ab} = h_{ab}-g_{ab} \ee
We use the expansion
\be \det h_{ab} = \det (g_{ab}+y_{ab}) = g \exp\sum_{n=1}^\infty \frac{(-1)^{n+1}}{n}\tr\left((g^{-1}y)^n\right) \ee
the square root of which is the action. Therefore the expansion of the NG action will include various powers of \(\tr\left((g^{-1}y)^n\right)\). We have to assume for this that \(g\) is invertible, which is not true at some points (in our case the folding points). Nevertheless, we proceed in writing such an expansion to see where it breaks down.

Expanding to quadratic order in \(y_{ab}\), we have
\be S_{NG} = - T\int d^2\sigma \sqrt{-g}\left(1+\frac12 \tr(g^{-1}y) -\frac14 \tr(g^{-1}yg^{-1}y) + \frac18(\tr(g^{-1}y))^2\right) +\ldots\ee
Note that \(y_{ab}\) includes both terms linear and quadratic in the fluctuations \(Y^\mu\), so the last equation contains some terms up to quartic order in \(Y\), but not all terms of that order. We keep only the quadratic terms in \(Y\) in the last equation to be consistent with the expansion. Terms linear in \(Y\) vanish because we assume the equation of motion \(\pa_a (\sqrt{-g} g^{ab}\pa_b \bar X^\mu) = 0\) holds.

The quadratic action for \(Y^\mu\), assuming now for simplicity a flat background, is
\begin{align} \nonumber S =& -T\int d^2\sigma \sqrt{-g} \left(1 - \frac12 g^{ab} \pa_a Y \cdot \pa_b Y + \frac12 (V^a \cdot \pa_a Y)^2\right) = \\ =&  -T \int d^2\sigma \sqrt{-g}\left(1 - \frac12 (g^{ab}\eta_{\mu\nu}-V^a_\mu V^b_\nu)\pa_a Y^\mu \pa_b Y^\nu \right)
\end{align}
where we have defined
\be V^{a,\mu} = g^{ab}\pa_b \bar X^\mu\ee
Fluctuations in directions transverse to the classical solution in the sense that they do not appear in the scalar product \(V^a \pa_a Y\) are free.

Without loss of generality we can make a choice of \(\tau\) and \(\sigma\) such that the classical solution obeys the conformal gauge constraint of \(g_{ab} =  e^{\varphi}\eta_{ab}\). Then the action is
\be S = -T\int d^2\sigma \left(e^{\varphi} - \frac12 (\eta^{ab}\eta_{\mu\nu} - e^{-\varphi}\eta_{\mu\lambda}\eta_{\nu\rho} \eta^{ac}\eta^{bd} \pa_c \bar X^\lambda \pa_d \bar X^\rho)\pa_a Y^\mu \pa_b Y^\nu\right) \label{flucaction} \ee
Now we see explicitly the factor of \(e^{-\varphi}\), which is proportional to the worldsheet curvature. When the curvature diverges, then some of the coefficients of the fluctuations in the semiclassical expansion will diverge as well. Then we will have divergences in the equations of motion which will have to be taken care of, and since the divergences are associated with the diff invariant curvature, they will be present in any gauge. On the other hand, fluctuations that are orthogonal to the classical solution do not know that they live on a worldsheet with a singular geometry.

In general the next step is to redefine the fluctuations by taking some linear transformation \(\tilde Y^\mu = f^\mu{}_\nu(\tau,\sigma)Y^\nu\) such that the \(\tilde Y\) have a canonical kinetic term. We do that next for the rotating string solution, and we show that the divergence remains for the redefined fluctuations.

\subsection{Expanding around rotating strings} \label{sec:exp_rot}
The general form of the solution rotating in the \(12\) plane which we will expand around is
\be \bar X^0 = \tau\,,\qquad \bar X^1 = \frac1\omega \cos(\omega\sigma+\phi) \cos(\omega\tau)\,,\qquad \bar X^2 = \frac1\omega \cos(\omega\sigma+\phi) \sin(\omega\tau) \ee
and with \(\bar X^i = 0\) for \(i\geq 3\). In the polar coordinates where \(X^1 = \rho \cos\theta\), \(X^2 = \rho \sin\theta\), the solution is
\be \rho = R(\sigma) = \frac1\omega\cos(\omega\sigma+\phi)\,, \qquad \theta = \omega\tau \ee

We now add fluctuations to the solution, taking \(X^\mu = \bar X^\mu + \delta X^\mu\). Of the \(D-2\) modes transverse to the string, one is in the plane of rotation and needs special treatment. This is the mode of fluctuations in the \(\theta\) direction which, once properly normalized, will be referred to as the planar mode. The modes transverse to the plane of rotation we call the transverse modes for short. The fluctuations in the time direction can be set to zero by the static gauge choice \(\tau = X^0\), or \(\delta X^0 = 0\). There is also the longitudinal mode \(\delta\rho\). Due to the reparametrization invariance in \(\sigma\), the bulk action for \(\delta\rho\) vanishes, and it only appears in boundary terms.

We expand the Nambu-Goto action to quadratic order in the fluctuations, as done in the last subsection.

The full bulk action for each of the transverse fluctuations (i.e. for \(f_t = \delta X^i\) with any \(i = 3,\ldots,D-1)\) is simply
\be S_t = T\int d^2\sigma \left(\frac12\dot f_t^2 - \frac12f_t^{\prime2}\right) \ee
These modes are free and are not affected by the rotation of the string.
The two modes in the plane of rotation have the action
\begin{align} S_p = T\int d^2\sigma \bigg[\frac12\dot f_p^2 -& \frac12f_p^{\prime2}-\frac{\omega^2}{\sin^2(\omega\sigma+\phi)}f_p^2 - \nonumber \\
-& \frac{\pa}{\pa\sigma}\left(\frac{\omega}{\sin[2(\omega\sigma+\phi)]}f_p^2-\frac{\omega\cot(\omega\sigma+\phi)}{2}f_r^2-\cos(\omega\sigma+\phi) f_r \dot f_p\right)\bigg] \label{eq:Sp} \end{align}
We have defined
\be f_r = \delta\rho\,,\qquad f_p = \frac{\cot(\omega\sigma+\phi)}{\omega}\delta\theta \ee
In the bulk the planar mode has a position dependent mass
\be \frac12 M_p^2(\sigma) = \frac{\omega^2}{\sin^2(\omega\sigma+\phi)} = \gamma(\sigma)^2\omega^2 \ee
which diverges at any point that moves at the speed of light, or equivalently for this solution, where the worldsheet curvature is infinite.

The action in eq. \ref{eq:Sp} also includes boundary terms. The coefficients of the first two boundary terms also diverge if the boundaries move at the speed of light, as in the case of an ordinary open string with Neumann boundary conditions.

In the following we will combine the action above with that for massive point particles, whose action is simply
\be S_m = -\sum_i m_i\int d\tau \sqrt{-\dot X^2}|_{\sigma=\sigma_i} \ee
at some set of given points \(\sigma_i\). These can be either the endpoints of the string, or in our case the folding points. If we expand this action in the fluctuations, we have
\be S_m = \sum_i \gamma_i m_i \int d\tau \left(\frac12\dot f_t^2 +\frac12\dot f_p^2+\frac12\dot f_r^2+\frac12\omega^2\gamma_i^2 f_r^2 + (\gamma_i+\frac1\gamma_i)\omega f_r \dot f_p\right) \ee
where \(\gamma_i = |\sin(\omega\sigma_i+\phi)|^{-1}\) now corresponds to a finite velocity of the points. The other boundary terms coming from the expansion of the NG action will also be made finite if the velocity is finite at the boundaries.

Regardless of the boundary terms, the bulk equation of motion for the planar mode is
\be f_p^\dprime - \ddot f_p - \frac{2\omega^2}{\sin^2(\omega\sigma+\phi)}f_p = 0  \ee
For the Fourier modes defined by
\be f_p(\tau,\sigma) = \alpha_0 f_0(\sigma) + i\sqrt{\frac\alp2}\sum_{n\neq0} \frac{\alpha_n}{\omega_n} e^{-i\omega\omega_n\tau}f_n(\sigma) \ee
We have to satisfy, in terms of \(x = \omega\sigma\)
\be f_n^\dprime(x) + \left(\omega_n^2-\frac{2}{\sin^2(x+\phi)}\right)f_n(x) = 0 \label{eq:planar}\ee
One way to solve the equation is to switch variables to \(y = \cos(x+\phi)\) and define \(f_n = (1-y^2)^{1/4}g_n(y)\). Then the equation for \(g_n\) is the Legendre equation,
\be (1-y^2)g_n^\dprime(y) - 2y g_n^\prime + \left(\omega_n^2-\frac14 - \frac{9}{4(1-y^2)}\right)g_n(y) = 0 \label{eq:Legendre} \ee
for which the general solution is given by the associated Legendre functions \(P_\nu^\mu(y)\) and \(Q_\nu^\mu(y)\), with \(\mu = 3/2\) and where \(\nu_n = \omega_n-\frac12\).

In terms of \(x\) the two linearly independent solutions to \ref{eq:planar} are in fact given by
\be s_1(x+\phi) = \frac{1}{|\sin(x+\phi)|}\left[(1-\omega_n)\cos(x+\phi)\sin\left(\omega_n(x+\phi)\right)-\omega_n\sin\left((1-\omega_n)(x+\phi)\right)\right] \label{eq:s1} \ee
\be s_2(x+\phi) = \frac{1}{|\sin(x+\phi)|}\left[(1-\omega_n)\cos(x+\phi)\cos\left(\omega_n(x+\phi)\right)+\omega_n\cos\left((1-\omega_n)(x+\phi)\right)\right] \label{eq:s2} \ee

For generic values of \(\omega_n\), the functions \(s_1\) and \(s_2\) both diverge at points where \(x+\phi\) is a multiple of \(\pi\). In the special case of \(\omega_n = n\), only \(s_2\) diverges at those points.

Because of the divergence at the folding points, the eigenmodes are not normalizable. To illustrate this, we can start with an ordinary Sturm-Liouville equation, of which eqs. \ref{eq:planar} and \ref{eq:Legendre} are particular cases,
\be \frac{d}{dx}\left(p(x)f_n^\prime \right) + \left(q(x) + \omega_n^2 r(x)\right) f_n = 0 \ee
In writing the equation, we already assume we can find a discrete spectrum of positive eigenvalues \(\lambda_n = \omega_n^2\). If we multiply the equation of motion by another solution, \(f_m(x)\) with \(\omega_n\neq\omega_m\), integrate over \(x\) in the interval \([a,b]\) where we wish to solve the equation, and then subtract the same equation with \(m\leftrightarrow n\), we obtain the relation
\be (\omega_n^2-\omega_m^2) \int_a^b dx r(x) f_m f_n = \int_a^b \left[f_n \frac{d}{dx}\left(p(x) f_m^\prime\right)-f_m \frac{d}{dx}\left(p(x) f_n^\prime\right)\right] \ee
If there are no divergences in the interval \([a,b]\) (including the boundary points), we can integrate the RHS by parts and get that
\be (\omega_n^2-\omega_m^2) \int_a^b dx r(x) f_m f_n = \left(f_n p(x) f_m^\prime-f_m p(x) f_n^\prime\right)|_a^b \ee
At the boundary points \(a\) and \(b\) there is typically a condition relating \(f\) to \(f^\prime\) so the above equation defines the appropriate inner product for two eigenfunctions on the interval \([a,b]\). In our examples we can show that the RHS is also proportional to \(\omega_n^2-\omega_m^2\), so one can divide by it and get an equation that will also be good for \(m=n\), fully defining the inner product of two eigenfunctions.

For the equation of motion of the fluctuations around the rotating string solution, \(p(x) = r(x) = 1\), but the eigenmodes themselves will be divergent at some points. The divergence will be either at the boundaries (Neumann open string) or at the folding points in rotating solutions that have them. We need to introduce some regulator in order to make all terms in the last equation finite, both the integral in the bulk and the boundary terms.

When we add a massive particle on the folding point, then we introduce a new boundary which contributes to the RHS. In general we can have some points \(x_i\) where the \(f_n\) are continuous but the derivatives \(f_n^\prime\) are not. We denote the discontinuity by \(\Delta f^\prime(x) \equiv \lim_{\epsilon\to0}(f^\prime(x+\epsilon)-f^\prime(x-\epsilon))\), and then we need to add to the last equation extra terms
\be (\omega_n^2-\omega_m^2) \int_a^b dx r(x) f_m f_n = \left(f_n p(x) f_m^\prime-f_m p(x) f_n^\prime\right)|_a^b - \sum_i \left(f_n p(x) \Delta f_m^\prime-f_m p(x) \Delta f_n^\prime\right)|_{x=x_i} \label{eq:ortho} \ee
With the mass terms both sides of the equation will be finite, with the boundary terms on the folding point offering a way to cancel the divergences of the integral when the mass is taken to zero. In this way, we have a well defined orthonormal set of eigenfunctions for any finite mass, and we have a controllable limit when the mass is zero which we use to define the solutions on the ordinary folded string with no masses.

\subsection{The intercept}
Our goal in the next few sections is to calculate the spectrum of fluctuations around the rotating solutions as well as the quantum correction, which is intercept of the Regge trajectories associated with the rotating string solutions.

The classical energy and angular momentum of the string define the trajectory as
\be J = J_{cl}(E) \ee
The function \(J_{cl}(E)\) can be simply \(\alp E^2\) as for the ordinary rotating open string, or the function defined parametrically by eqs. \ref{eq:classical_E}-\ref{eq:classical_J} for the string with massive endpoints.
In all cases we can define the intercept as the correction to this classical relation, namely as the expectation value
\be a \equiv \langle J - J_{cl}(E) \rangle \ee
In terms of the fluctuations of the string, at quadratic order in the fluctuations, it is
\be a = \langle \delta J - \frac{\pa J_{cl}}{\pa E}\delta E\rangle =\langle \delta J - \frac{1}{\omega}\delta E \rangle = -\frac{1}{\omega}\langle H_{ws}\rangle \ee
where \(\delta E\) and \(\delta J\) are the contributions to the energy and angular momentum of the quantum fluctuations, and the equality \(\frac{\pa J_{cl}}{\pa E} = \frac1\omega\) can be shown using the full expressions for \(E\) and \(J\) (including in the case of a string coupled to massive particles). In the last step, this combination of \(\delta J\) and \(\delta E\) when computed can be shown to be proportional to the worldsheet Hamiltonian of the fluctuations, \(H_{ws}\). Therefore, the intercept is given simply by the expectation value of the Hamiltonian, which in turn will be a normal ordering constant as in the Neumann open string.

For non-critical strings part of the correction will be another contribution from the Polchinski-Strominger term that is added to the effective string action in for \(D\neq26\). In sections \ref{sec:quantc} and \ref{sec:quanto} we quantize the fluctuations around closed and open rotating strings with folds, respectively. The analysis in those sections applies to both critical and non-critical strings. In section \ref{sec:PS} we calculate the PS term contribution to the intercept for both systems when the strings are in non-critical dimensions.

Throughout the paper we use only the quadratic approximation as detailed above, neglecting higher order terms. The reason is that formally we expand in \(\ell_s/L\) where \(\ell_s = \sqrt{\alp}\) is the characteristic string length, and \(L\) the length of the string in the classical solution. The divergence associated with the fold is a property of the zeroth order classical solution, so we expect to see its effects at any (finite) order of this expansion. As a result, many terms in the action will come with coefficients which are divergent at the folding points. The assumption that we make when we truncate the expansion at quadratic order is that when we calculate the physical quantities such as the spectrum of fluctuations or the intercept, then the corrections from higher order terms in the fluctuations, once properly renormalized to eliminate the infinities from the folding points, will be of the expected order in terms of the \(1/L\) expansion and therefore negligible. This is because the infinities are associated with a point on the string and therefore independent of the length. 

\section{Quantizing the folded closed string} \label{sec:quantc}
\subsection{Expanding around rotating closed string}
In this section we quantize the fluctuations around the rotating folded closed string solution which we wrote in section \ref{sec:classical},
\be X^0 = \tau \qquad X^1 = \frac1\omega\cos(\omega\sigma)\cos(\omega\tau) \qquad X^2 = \frac1\omega\cos(\omega\sigma)\sin(\omega\tau) \ee
where \(\sigma \in [-\ell,\ell]\) and \(\omega = \pi/\ell\).

The fluctuations transverse to the plane of rotation are free, and behave as in ordinary non-rotating strings. For the planar mode the solutions to the equation of motion (eq. \ref{eq:planar} with \(\phi=0\)) are
\be s_1(x) = \frac{1}{|\sin x|}\left[(1-\omega_n)\cos x\sin\left(\omega_n x\right)-\omega_n\sin\left((1-\omega_n)x\right)\right] \ee
\be s_2(x) = \frac{1}{|\sin x|}\left[(1-\omega_n)\cos x\cos\left(\omega_n x\right)+\omega_n\cos\left((1-\omega_n) x\right)\right] \ee
where \(x = \pil\sigma\). Since the only boundary condition is periodicity, \(f_p(\tau,-\ell) = f_p(\tau,\ell)\), then the allowed eigenfrequencies are integers, \(\omega_n = n\).

\begin{figure}[h!] \centering
	\includegraphics[width=0.6\textwidth]{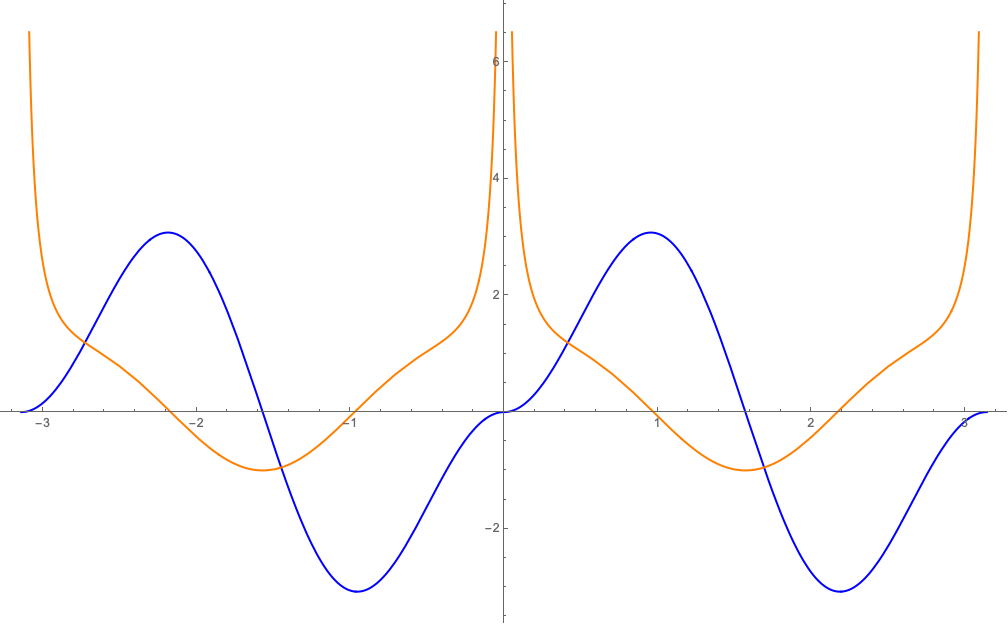}
	\caption{\label{fig:closedmodes} Eigenmodes of the planar fluctuations on the closed rotating string, with \(n=3\). The even mode diverges at the folding points \(0\) and \(\pi\), the odd modes vanish there.}
\end{figure}

We have two independent modes at each \(n\), the equivalent of left and right moving modes in the non-rotating case. The modes \(s_1\) are odd under \(x\to -x\) while \(s_2\) are even.
Since the folded string solution consists of two segments of the string sitting on top of each other, the even modes are those where the two segments move together, while the odd modes are those where they move away from each other. The even modes are identical to the fluctuations found on an open string with Neumann boundary conditions, and they are the modes that diverge at the folding point. The odd modes vanish at the folding point and can be likened to the modes on open strings with Dirichlet boundary conditions. We plot an example in figure \ref{fig:closedmodes}

As explained in section \ref{sec:exp_rot} the divergent modes are not normalizable before the addition of a regulating mass term.

\subsection{Closed string with massive folds} \label{sec:closed_masses}
We place two masses, \(m_0\) and \(m_\ell\) on the two folding points of the rotating closed string solution, \(\sigma=0\) and \(\ell\) (equivalently \(-\ell\)). Eventually we will take the massless limit to return to the ordinary closed string, but we solve the system for any value of the masses. The massless limit will not depend on the ratio of \(m_0/m_\ell\).

Note that for given values of the masses, we should expand around a solution with \(m_i L \gg 1\) and \(T L/m_i \gg 1\), where \(L\) is the total length of the string, otherwise we need to include higher order terms in the fluctuations \cite{Sonnenschein:2018aqf}.

\subsubsection{Classical rotating solution with massive folding points}
We define the classical solution piecewise as
\be X^0 = \tau \qquad \rho = \begin{cases} \frac1\omega \cos(\omega\sigma-\phi) \qquad -\ell<\sigma<0\\ \frac1\omega \cos(\omega\sigma+\phi) \qquad 0<\sigma<\ell \end{cases} \qquad \theta = \omega\tau \ee
where \(\rho\) and \(\theta\) are the polar coordinates in the 12 plane. The solution is such that it and its time derivatives are continuous everywhere, and in particular \(X(\tau,-\ell) = X(\tau,\ell)\).

However, there is a discontinuity in the derivative \(X^\prime\) at the folding points \(\sigma = 0\) and \(\ell\), and the jump in the derivative is determined by the equations of motion of the massive particles on those points:
\be T (X^{\prime\mu}|_{\sigma=0^-}-X^{\prime\mu}|_{\sigma=0^+}) + m_0 \pa_\tau\left(\frac{\dot X^\mu}{\sqrt{-\dot X^2}}\right) = 0 \ee
\be T (X^{\prime\mu}|_{\sigma=-\ell^+}-X^{\prime\mu}|_{\sigma=\ell^-}) - m_\ell \pa_\tau\left(\frac{\dot X^\mu}{\sqrt{-\dot X^2}}\right) = 0 \ee
From here we have two conditions determining \(\phi\) and \(\omega\). We define the useful parameter \(\delta\equiv \omega\ell\), then
\be \frac{\sin\phi}{\cot\phi} = \frac{m_0\omega}{2T}\,, \qquad \frac{m_\ell \omega}{2T}=-\frac{\sin(\delta+\phi)}{\cot(\delta+\phi)}\ee
are the two boundary conditions. These are the same equations as for a rotating open string solution with massive endpoints (eq. \ref{eq:bdc}), written in terms of different parameters.

We pick the solutions in the range where \(0<\phi<\pi/2\), \(\pi/2<\delta+\phi<\pi\). The velocities at which the folding points move are now
\be \beta_0 = \cos\phi \qquad \beta_\ell = -\cos(\delta+\phi) \ee
\be \gamma_0 = \frac{1}{\sin\phi} \qquad \gamma_\ell = \frac1{\sin(\delta+\phi)} \ee
Going the other way, \(\phi = \arccos\beta_0\) while \(\delta = \arcsin\beta_0+\arcsin\beta_\ell\). In the massless limit we return to \(\phi=0\) and \(\delta=\pi\) where the folds move at the speed of light. The special case of equal masses has \(\delta = \pi-2\phi\).

The total length of the string is given by
\be L = \frac{2}{\omega}(\beta_0+\beta_\ell) \ee
while its energy and angular momentum are
\be E = \gamma_0 m_0 + \gamma_\ell m_\ell + TL\frac{\arcsin\beta_0+\arcsin\beta_\ell}{\beta_0+\beta_\ell}  \label{eq:classical_E_c}
\ee
\be J = \frac12\frac{\gamma_0 m_0 \beta_0^2+\gamma_\ell m_\ell \beta_\ell^2}{\beta_0+\beta_\ell}L + \frac14 \frac{\gamma_0\gamma_\ell(\arcsin\beta_0+\arcsin\beta_\ell)-\gamma_0\beta_\ell-\gamma_\ell\beta_0}{\gamma_0\gamma_\ell(\beta_0+\beta_\ell)^2}TL^2
\ee
The corresponding Regge trajectory is
\be J = \frac{\alp}{2}E^2\left(1-\frac{4\sqrt\pi}{3}(m_0^{3/2}+m_\ell^{3/2})E^{-3/2} + \ldots \right) \ee
Classically this string is completely equivalent to an open string with massive endpoints written in section \ref{sec:classicalMass}, with the string tension effectively doubled for the folded string. In the remainder of the section we analyze the quantum fluctuations.

\subsubsection{Transverse fluctuations} \label{sec:ct}
The transverse modes do not require special treatment in the presence of the folds, but since we add Lorentz invariant mass terms to the action the masses will affect all the modes. In this section we show the procedure of adding the mass terms and taking the massless limit to recover the expected result for the contribution to the intercept of each transverse mode, which are two contributions of \(\frac{1}{24}\), one each from the would be left and right moving modes.

The action for each of the transverse fluctuation modes to quadratic order is
\be S_t = T\int d^2\sigma (\frac12\dot f_t^2 - \frac12 f_t^{\prime2}) + \gamma_0 m_0 \int d\tau \frac12 \dot f_t^2|_{\sigma=0} + \gamma_\ell m_\ell \int d\tau \frac12 \dot f_t^2|_{\sigma=\ell} \ee
There are \(D-3\) such modes in the directions \(X^i\) with \(i=3,\ldots,D-1\), but we omit the spatial index for the rest of this section, since the modes are identical and independent.

The ``boundary'' conditions are continuity at \(\sigma = 0\), periodicity in the sense of continuity at \(\pm\ell\), \(f_n(-\ell)=f_n(\ell)\), and the jump conditions at the folds,
\be T(f_n^\prime|_{0^+} - f_n^\prime|_{0^-}) + \omega_n^2 \gamma_0 m_0 \omega f_n = 0, \qquad x=0 \ee
\be T(f_n^\prime|_{-\ell^+} - f_n^\prime|_{\ell^-}) + \omega_n^2 \gamma_\ell m_\ell \omega f_n = 0, \qquad x=\omega\ell = \delta\ee
One can start from the most general ansatz for the solution to the bulk equation,
\be f_n(x) = \begin{cases} c_1 \cos(\omega_n x) + c_2\sin(\omega_n x) & -\delta \leq x \leq 0 \\ c_3 \cos(\omega_n x) + c_4\sin(\omega_n x)  & 0\leq x \leq \delta \end{cases} \ee
and find that we have two independent types of solutions, which are the even and odd modes. Each of them will have a different set of eigenfrequencies, so the general mode expansion we write is
\be f_t(\tau,\sigma) = i\sqrt{\frac\alp2}\sum_{n\neq0}\left(\frac{\alpha_n}{\omega_n} e^{-i\omega\omega_n\tau}f^+_n(\omega\sigma) + \frac{\tilde\alpha_n}{\tilde \omega_n}e^{-i\omega\tilde\omega_n\tau}f^-_n(\omega\sigma)\right) \label{eq:modexpclosed}\ee

The even modes satisfying all the conditions take the form (up to an overall normalization constant)
\be   f^+_n(x) = \begin{cases}    \cos\phi\cos( \omega_n x) +  \omega_n\sin\phi\sin( \omega_n x) & -\delta \leq x \leq 0 \\    \cos\phi\cos( \omega_n x) -  \omega_n\sin\phi\sin( \omega_n x) & 0\leq x \leq \delta \end{cases} \ee
where the eigenfrequencies are the solutions of
\be [\tan\phi - \tan(\delta+\phi)]\omega_n \cos(\omega_n \delta) + [1+\tan\phi\tan(\delta+\phi)\omega_n^2]\sin(\omega_n \delta) = 0 \ee
Or, in terms of the velocities of the folding points,
\be \left(\frac{1}{\gamma_0 \gamma_\ell}\omega_n^2 - \beta_0\beta_\ell\right)\sin(\omega_n \delta) - \omega_n \left(\frac{\beta_0}{\gamma_\ell}+\frac{\beta_\ell}{\gamma_0}\right)\cos(\omega_n \delta) = 0 
\label{eq:ftce}\ee
This is exactly the same equation as for the string with massive endpoints found in \cite{Sonnenschein:2018aqf} (eq. 8.2 there). At small masses (large \(\gamma\)) we can write an approximate solution to the equation,
\be \omega_n = n + \frac{n^3-n}{3\pi}\left(\frac{1}{\gamma_0^3}+\frac{1}{\gamma_\ell^3}\right) + \mathcal{O}(\gamma^{-5}) \label{eq:wtce}\ee
Note that the coefficients in the expansion grow with \(n\), such that the expansion parameter is  effectively \(n/\gamma\).

On the closed string we also have the odd modes, which are not present in the rotating open string. Here they are simply given by
\be f^-_n(x) = \sin(\tilde\omega_n x) \ee
and the eigenfrequencies are
\be   \sin(\tilde\omega_n \delta)  = 0 \qquad \Rightarrow \qquad \tilde\omega_n = \frac{\pi}{\delta}n
\label{eq:ftco}\ee
The odd modes vanish at the folding points, their derivatives are continuous there and in general are unaffected by the mass except for the overall factor of \(\delta\) in the eigenfrequencies.

Per equation \ref{eq:ortho}, all the eigenmodes should satisfy
\be (\omega_n^2-\omega_m^2) \int_{-\delta}^{\delta} dx f_m f_n = -\left(f_n \Delta f_m^\prime- f_m \Delta f_n^\prime\right)|_{x=0} -\left(f_n \Delta f_m^\prime- f_m \Delta f_n^\prime\right)|_{x=\delta} \ee
We can use the boundary conditions to simplify this. For the even modes
\be  \int_{-\delta}^{\delta} dx f^+_m f^+_n + \frac{\gamma_0 m_0 \omega}{T}f^+_m f^+_n|_{x=0} + \frac{\gamma_\ell m_\ell \omega}{T}f^+_m f^+_n|_{x=\delta} =2\pi (\delta_{m-n} + \delta_{m+n}) \label{eq:525} \ee
For the odd modes the boundary terms are trivial, so
\be  \int_{-\delta}^{\delta} dx f^-_m f^-_n  =2\pi (\delta_{m-n} + \delta_{m+n}) \label{eq:526} \ee
Between them, even and odd modes are obviously orthogonal.

The canonical quantization is done by imposing
\be [\alpha_m,\alpha_n] = \omega_{m}\delta_{m+n} \qquad [\tilde\alpha_m,\tilde\alpha_n]= \tilde\omega_m \delta_{m+n} \qquad [\alpha_m,\tilde \alpha_n] = 0 \label{eq:commalpha}\ee
which guaranties that
\be [f_t(\tau,\sigma),\pi_t(\tau,\sigma^\prime)] = i\delta(\sigma-\sigma^\prime) \ee
where \(\pi_t\) is the canonical momentum conjugate to \(f_t\).

The worldsheet Hamiltonian is
\be H_t = T\int_{-\ell}^\ell d\sigma (\frac12 \dot f_t^2 + \frac12 f_t^{\prime2}) + \frac12\gamma_0 m_0 \dot f_t^2|_{\sigma=0} + \frac12\gamma_\ell m_\ell \dot f_t^2|_{\sigma=\ell} \ee
Upon inserting the mode expansion and using eqs. \ref{eq:525} and \ref{eq:526}, it becomes
\be H_t = \frac{\omega}{2}\sum_{n\neq 0}(\alpha_{-n}\alpha_n + \tilde \alpha_{-n}\tilde \alpha_n) \ee
After normal ordering,
\be \frac1\omega H_t = \sum_{n=1}^\infty (\alpha_{-n}\alpha_n + \tilde\alpha_{-n}\tilde\alpha_n) - (A + \tilde A) \ee
The normal ordering constant is the intercept,
\be a_t = -\frac{1}{\omega}\langle H_t \rangle = A + \tilde A = -\frac12\sum_{n=1}^\infty \omega_n -\frac12\sum_{n=1}^\infty \tilde \omega_n \ee

The task now is to compute and renormalize the divergent sum over the eigenfrequencies. There are two methods that we can use. First is the Zeta function regularization, and the in the second method we first convert the sum into a contour integral, as first proposed in \cite{Lambiase:1995st}. The calculation and renormalization done are almost identical to the one performed for the open string with massive endpoints in \cite{Sonnenschein:2018aqf}. We repeat the key points below.

The equations from which the eigenfrequencies are determined (eqs. \ref{eq:ftce} and \ref{eq:ftco}) are of the form \(g(\omega_n) = 0\) for some function \(g\). If the function \(g(\omega)\) when defined in the complex plane has only simple zeroes at \(\omega=\omega_n\) and no poles, then the integral
\be \frac{1}{2\pi i}\oint_{\mathcal{C}} dz z \frac{d}{dz}\log g(z) = \sum_{\omega_n \in \mathcal C} \omega_n \ee
gives the sum over all zeroes inside the contour \(\mathcal C\).

\begin{figure}[ht!] \centering
\includegraphics[width=0.37\textwidth]{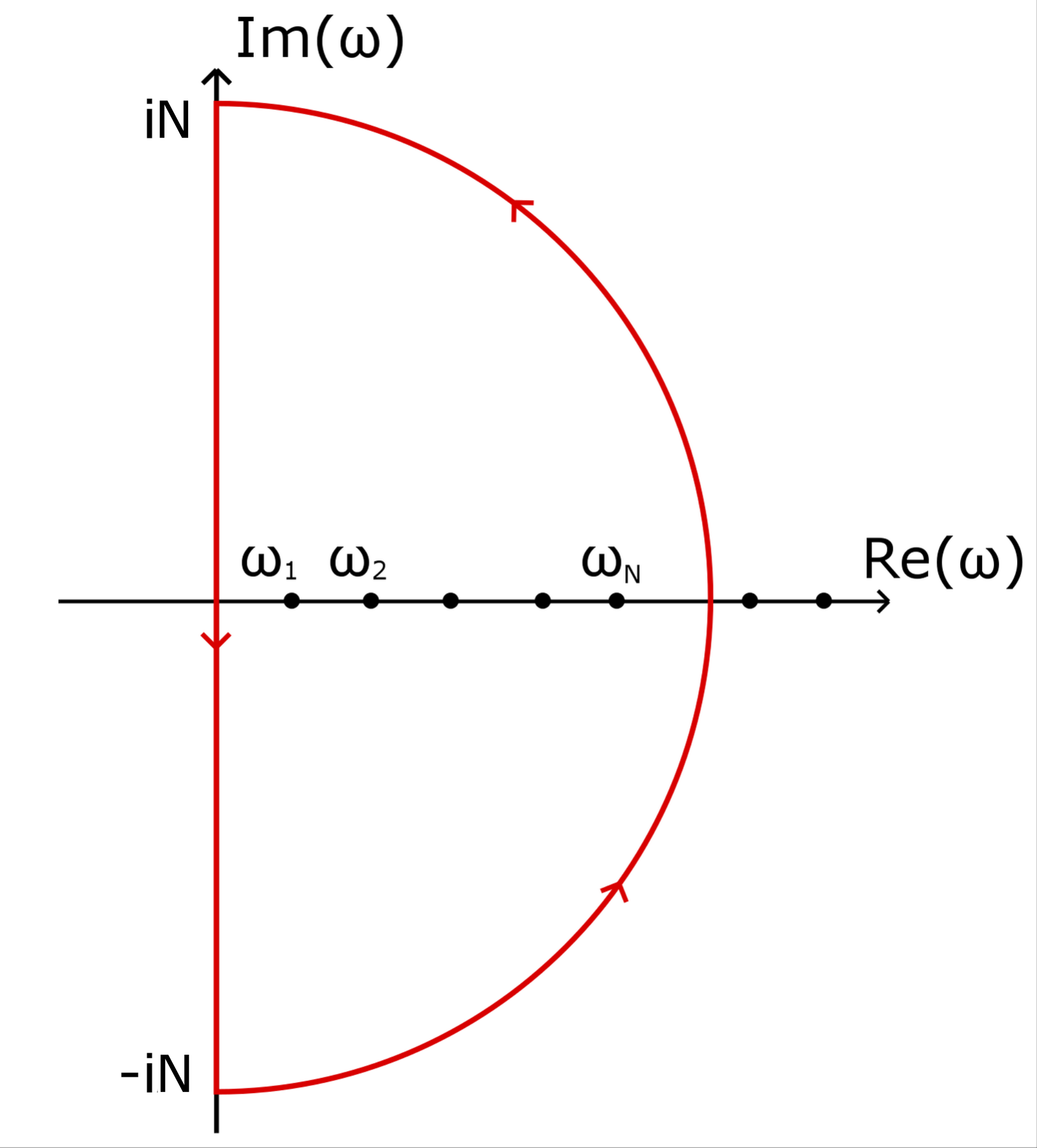}
\caption{\label{fig:contour} Semicircular contour over which we carry out the integral that will sum over the eigenfrequencies on the positive real axis.}
\end{figure}

In order to do the renormalization procedure, we compute the (regularized) Casimir energy. The dimensionful eigenfrequencies are \(\Omega_n = \omega \omega_n\), and we sum up all the energies to some cutoff scale \(\Lambda\), which we also write as \(\Lambda = \omega N\). Then we convert the sum into a contour integral
\be E_C(\Lambda) \equiv \frac\omega2 \sum_{n=1}^N \omega_n = \frac{\omega}{4\pi i}\oint_{\mathcal C(N)} dz z\frac{d}{dz}\log g(z) \ee
The contour is the closed semicircle with radius \(N\) as depicted in figure \ref{fig:contour}. 

We compute the contour integral explicitly at large \(\Lambda\) (\(N\)), to see the form of the divergences in the \(\Lambda\to\infty\) limit for which we then offer a prescription for renormalization.

Starting with the odd modes, for which \(g(z) = \sin(\delta z)\), we can show by explicit calculation of the contour integral that the Casimir energy associated with these modes is
\be E_C^-(\Lambda) \equiv \frac\omega2\sum_{n=1}^N \tilde\omega_n = \left(-\frac{\Lambda^2\delta}{4\pi\omega} - \frac{\pi\omega}{24\delta}\right) + \left(\frac{\Lambda^2\delta}{2\pi\omega}\right)\ee
with the first two terms on the RHS coming from the integral on the imaginary axis, and the third term from the integral on the semicircle (see appendix \ref{app:contour} for a more detailed calculation). In total
\be E_C^-(\Lambda) = \frac{\Lambda^2}{4\pi}\ell - \frac{\pi}{24\ell} \ee
The quadratic divergence is proportional to \(\ell\), which appears in the expression for the classical energy of the rotating string, eq. \ref{eq:classical_E_c}, which states that \(E_{cl} = \gamma_0 m_0 + \gamma_\ell m_\ell + T \ell\). Therefore the divergence can be subtracted by an appropriate redefinition of the string tension. One can also think of it as subtracting the contribution from an infinitely long string from the Casimir force \(F_C = -\pa E_C /\pa \ell\). After the subtraction of the quadratically divergent part, the term that goes like \(1/\ell\) gives precisely the intercept, and the result matches that of the Zeta function regularization.
\be \tilde A = -\frac12\sum_{n=1}^\infty \frac{\pi}{\delta}n = \frac\pi\delta \times \frac{1}{24} \ee

For the even modes, the calculation and the result are the same as for the string with massive endpoints. One can do the contour integral with a function \(f(z)\) matching eq. \ref{eq:ftce}. In \cite{Sonnenschein:2018aqf} we have shown that the divergent parts are given by\footnote{In fact here we give the generalization to two different masses of eq. 4.100 in \cite{Sonnenschein:2018aqf}.}
\be E_C^+(\Lambda) = \frac{\Lambda^2}{2\pi}\ell + \frac{1}{\pi}\left(\frac{2T}{\gamma_0 m_0}+\frac{2T}{\gamma_\ell m_\ell}\right)\frac{\log\left(\frac{\gamma_0 m_0 \Lambda}{2T}\right)\gamma_0 \beta_0-\log\left(\frac{\gamma_\ell m_\ell \Lambda}{2T}\right)\gamma_\ell \beta_\ell}{\gamma_0\beta_0-\gamma_\ell \beta_\ell} + (\mathrm{finite}) \ee
Now in addition to the quadratic divergence which we absorb into the tension, there are logarithmic divergences which we renormalize by redefinition of the masses. Note that the masses always appear as \(\gamma m\), as they do in the action.

We can write the divergent parts of the full contour integral as another contour integral over an asymptotic form of \(f(z)\), and subtract them in that form, such that the finite intercept is given by the difference. The exact answer for the regularized sum of the eigenfrequencies is given then in integral form as
\be A = -\frac{1}{2\pi}\int_0^\infty dy \log\left(1-e^{-2\delta y}\frac{(y-\gamma_0\beta_0^2)(y-\gamma_\ell\beta_\ell^2)}{(y+\gamma_0\beta_0^2)(y+\gamma_\ell\beta_\ell^2)}\right) \label{eq:atcei} \ee
When the masses go to zero this reduces to
\be A \to -\frac{1}{2\pi} \int_0^\infty dy \log(1-e^{-2\pi y}) = \frac{1}{24} \ee
As shown in appendix \ref{app:Zeta}, when we expand the integral in \(\gamma^{-1}\) when we go to small masses, we find that the corrections match exactly what we can get from expanding \(\omega_n\) as in eq. \ref{eq:wtce} and performing Zeta function regularization at each order. The results are
\be A = \frac{1}{24} - \frac{11}{720\pi}(\epsilon_1^3+\epsilon_2^3) + \ldots \ee
\be \tilde A = \frac{1}{24}\left(1+\frac{1}{\pi}(\epsilon_1+\epsilon_2)+\frac{1}{\pi^2}(\epsilon_1+\epsilon_2)^2+\frac{1}{\pi^3}(\epsilon_1+\epsilon_2)^3+\frac{1}{6\pi}(\epsilon_1^3+\epsilon_2^3)\right) \ee
where \(\epsilon_1 \equiv 1/\gamma_0\) and \(\epsilon_2\equiv 1/\gamma_\ell\). Adding the two contributions,
\be a_t = A+\tilde A = \frac{1}{12}\left(1+\frac{1}{2\pi}(\epsilon_1+\epsilon_2)+\frac{1}{2\pi^2}(\epsilon_1+\epsilon_2)^3+\frac{1}{2\pi^3}(\epsilon_1+\epsilon_2)^3-\frac{1}{10\pi}(\epsilon_1^3+\epsilon_2^3)\right) \label{eq:atce} \ee
If the endpoint masses are equal \(\epsilon_1 = \epsilon_2 = \frac{1}{\gamma}\), then
\be a_t = \frac{1}{12}\left(1+\frac{1}{\pi}\frac1\gamma+\frac{2}{\pi^2}\frac{1}{\gamma^2}+\frac{20-\pi^2}{5\pi^3}\frac{1}{\gamma^3}\right) \ee
At the massless limit, we always recover the result \(a_t = \frac{1}{12}\). It is noteworthy that for small masses, the leading order correction has a positive sign, increasing the intercept. This comes from the odd modes, not present on the open string, whose intercept increases as one reduces the endpoint velocity. At small masses this correction is the dominant one, starting at order \(\gamma^{-1}\) while the even, open string modes only receive corrections at order \(\gamma^{-3}\).

\subsubsection{Planar fluctuations}
The planar fluctuations around the rotating solution with masses are defined as
\be f_p = \begin{cases} -\frac{\cot(\omega\sigma-\phi)}{\omega}\delta\theta & -\ell\leq \sigma \leq 0 \\ \frac{\cot(\omega\sigma+\phi)}{\omega}\delta\theta & 0\leq \sigma \leq \ell \end{cases} \ee
With the relative minus sign this definition is continuous at both \(0\) and \(\pm\ell\).
The bulk equations of motion for the fluctuations will be
\be f_p^\dprime - \ddot f_p - \frac{2\omega^2}{\sin^2(\omega\sigma-\phi)}f_p = 0\,,\qquad -\ell<\sigma<0 \ee
\be f_p^\dprime - \ddot f_p - \frac{2\omega^2}{\sin^2(\omega\sigma+\phi)}f_p = 0\,,\qquad 0<\sigma<\ell \ee
We write a mode expansion as in eq. \ref{eq:modexpclosed}, and in general the eigenmodes will be given by
\be f_n(\sigma) = \begin{cases} c_1 s_1(x-\phi) + c_2 s_2(x-\phi) & -\delta \leq x \leq 0 \\ c_3 s_1(x+\phi) + c_4 s_2(x+\phi) & 0 < x \leq \delta \end{cases} \label{eq:fpcgen}\ee
where \(x=\omega\sigma\) and \(s_1\), \(s_2\) are the solutions to the bulk equation given in eqs. \ref{eq:s1} and \ref{eq:s2}.

At \(\sigma = 0\) and \(\ell\) the action reads
\be S_{m_i} = \gamma_i m_i\int d\tau \left(\frac12\dot f_p^2 +\frac12\mu_{p,i}^2 f_p^2 + \frac12\dot f_r^2 + \frac12 f_r^2 + c_i f_r \dot f_p \right) \ee
with
\be \mu_{p,i}^2 = \gamma_i^2\omega^2 \qquad \mu_{r,i}^2=(2\gamma_i^2-1)\omega^2 \qquad c_i = 2\gamma_i\omega \ee

The variation of the action w.r.t. \(f_p\) yields the bulk equations of motion written above as well as the boundary conditions
\be T \Delta f_p^\prime + \gamma_i m_i (- \ddot f_p + \mu_{p,i}^2 f_p - c_i\dot f_r) = 0, \qquad \sigma = 0,\ell \ee
At the folding points there must be a discontinuity in \(f_p^\prime\), while \(f_p\) itself is continuous there.
The variation of the action on the folds w.r.t. \(f_r\) results in
\be -\ddot f_r + \mu_{r,i}^2 f_r + c_i \dot f_p = 0\,,\qquad \sigma = 0,\ell\ee

The mode expansion for \(f_p\) is exactly of the same form as in eq. \ref{eq:modexpclosed}, and again we will have even modes and odd modes. The mode expansion for \(f_r\), which lives only at the two folding points, will have the same set of eigenfrequencies, and we write the mode expansion as
\be f_r =  i\sqrt{\frac\alp2}\sum_{n} \frac{\alpha_n}{\omega_n} f_r^{(n)} e^{-i\omega \omega_n\tau} \ee
where the coefficients \(f_r^{(n)}\) are just constants (defined separately at each boundary point). The \(\alpha_n\) in the expansion are the oscillators corresponding to the even modes of \(f_p\). The radial mode adds no independent oscillators, and the odd modes of \(f_p\) vanish at the folding points and do not contribute to \(f_r\) by the equation of motion above.

We solve the equations for \(f_r\) at each of the boundary points to get a boundary condition only involving the modes of \(f_p\). Then we can substitute
\be f_r^{(n)} = \frac{i\omega \omega_n c_i}{\omega^2\omega_n^2+\mu_{r,i}^2}f^+_n(\sigma_i) \ee
where \(f^+_n(\sigma)\) are the even eigenmodes of the planar mode \(f_p\). 

The boundary conditions for the planar mode at the folding points are then
\begin{align} &f_n^\prime|_{0^+} - f_n^\prime|_{0^-} + \frac{\gamma_0 m_0}{T\omega}(\omega^2\omega_n^2+\mu_{p0}^2 - \frac{c_0^2\omega^2\omega_n^2}{\omega^2\omega_n^2+\mu_{r0}^2})f_n = 0\,, \qquad& x=0 \\
&f_n^\prime|_{-\ell^+} - f_n^\prime|_{\ell^-} + \frac{\gamma_\ell m_\ell}{T\omega}(\omega^2\omega_n^2+\mu_{p\ell}^2 - \frac{c_\ell^2\omega^2\omega_n^2}{\omega^2\omega_n^2+\mu_{r\ell}^2})f_n = 0\,, \qquad& x=\omega\ell=\delta \end{align}
We require also continuity of \(f_n\) at the points \(x = 0\) and \(x=\delta\), in the sense that \(f_n(-\ell)=f_n(\ell)\).

The odd modes have \(c_1 = c_3\), \(c_2 = -c_4\) (in the notation of eq. \ref{eq:fpcgen}). Then, the eigenfrequencies are the solution of the equation
\begin{align} \nonumber &[\tilde\omega_n^2\sin(\delta+\phi)\sin\phi + \cos(\delta+\phi)\cos(\phi)]\sin(\tilde\omega_n \delta) + \\ &+ \tilde\omega_n [\cos(\delta+\phi)\sin\phi - \cos\phi\sin(\delta+\phi)]\cos(\tilde\omega_n \delta) = 0 \end{align}
In terms of the velocities of the folding points the odd modes equation is
\be \left(\frac{1}{\gamma_0 \gamma_\ell}\tilde\omega_n^2-\beta_0 \beta_\ell\right)\sin(\tilde\omega_n \delta) - \tilde\omega_n \left(\frac{\beta_0}{\gamma_\ell}+\frac{\beta_\ell}{\gamma_0}\right)\cos(\tilde\omega_n \delta) = 0 \label{eq:fpco} \ee
and \(\delta = \arcsin\beta_0 + \arcsin\beta_\ell\).  Surprisingly, we find that this is the exact same equation as for the even transverse modes (eq. \ref{eq:ftce}).

For the even modes \(c_2 = c_4\), \(c_1 = -c_3\), and the eigenfrequencies are the solutions of
\begin{align} \nonumber&\left[4 \omega _n^2 \sin (2 \phi ) \sin (2 (\delta +\phi ))+\left(-\omega _n^2+\left(\omega
   _n^2+1\right) \cos (2 \phi )+3\right) \left(\left(\omega _n^2+1\right) \cos (2 (\delta
   +\phi ))-\omega _n^2+3\right)\right]\sin(\omega_n \delta) + \\
   &+ \left[-4 \sin (\delta ) \omega _n \left(4 \cos (\phi ) \cos (\delta +\phi )+2 \left(\omega
   _n^2-1\right) \sin (\phi ) \sin (\delta +\phi )\right)\right]\cos(\omega_n \delta) = 0
\end{align}
which can be written as   
\begin{align}\nonumber &\left[\frac{1}{\gamma_0^2\gamma_\ell^2}\omega_n^4 -2 \left(1-\beta_0^2\beta_\ell^2 + \frac{2\beta_0\beta_\ell}{\gamma_0\gamma_\ell}\right)\omega_n^2 + (1+\beta_0^2)(1+\beta_\ell^2)\right]\sin(\omega_n \delta)+\\&+2\omega_n \left(\frac{\beta_0}{\gamma_\ell}+\frac{\beta_\ell}{\gamma_0}\right)\left(\frac{1-\omega_n^2}{\gamma_0\gamma_\ell}+2\beta_0\beta_\ell\right)\cos(\omega_n \delta) = 0 \label{eq:fpce}\end{align}
This is also the equation found for the planar mode on an open string with endpoint masses.\footnote{In fact this is a generalization of the result of \cite{Sonnenschein:2018aqf} (eq. 5.22 there) to the case of two different masses.}

The calculation now proceeds in the same way as for the transverse modes. The eigenmodes satisfy the equations
\begin{multline}
\int_{-\delta}^{\delta} dx f^s_m f^{s^\prime}_n + \frac{\gamma_0 m_0 \omega}{T}\left(1-\frac{c_0^2\mu_{r0}^2}{(\omega^2\omega_m^2+\mu_{r0}^2)(\omega^2\omega_n^2+\mu_{r0}^2)}\right)f^s_m f^{s^\prime}_n|_{x=0}  \\+ \frac{\gamma_\ell m_\ell \omega}{T}\left(1-\frac{c_\ell^2\mu_{r\ell}^2}{(\omega^2\omega_m^2+\mu_{r\ell}^2)(\omega^2\omega_n^2+\mu_{r\ell}^2)}\right)f^s_m f^{s^\prime}_n|_{x=\delta} = 2\pi (\delta_{m-n} + \delta_{m+n})]\delta^{ss^\prime} \label{eq:closed_planar_norm}
\end{multline}
where the indices \(s\), \(s^\prime\) = \((+,-)\), now signify the even and odd modes. The odd modes do not require special treatment as their boundary terms vanish, but the even eigenmodes, which diverge on the folding points at the massless limit, are now normalizable in the way defined by this equation. The extra boundary terms are exactly what is needed to cancel out the divergent parts, as we can see by evaluating the integral and the boundary terms for small masses,\footnote{We write the functions up to an overall mass independent normalization constant.}
\be \int_{-\ell}^{\ell} d\sigma f_n^2 = 2(\gamma_0+\gamma_\ell) + (n^2-1)\pi + \mathcal O(\frac{1}{\gamma}) \label{eq:nrmff2}\ee
\be \frac{\gamma_0 m_0 \omega}{T}\left(1-\frac{c_0^2\mu_{r0}^2}{(\omega^2\omega_n^2+\mu_{r0}^2)^2}\right)f_n^2|_{\sigma=0} = -2\gamma_0 + \mathcal O(\frac{1}{\gamma_0}) \ee
\be \frac{\gamma_\ell m_\ell \omega}{T}\left(1-\frac{c_\ell^2\mu_{r\ell}^2}{(\omega^2\omega_n^2+\mu_{r\ell}^2)^2}\right)f_n^2 |_{\sigma=\ell} = -2\gamma_\ell + \mathcal O(\frac{1}{\gamma_\ell}) \ee
So the sum of integral plus boundary terms is finite when we take either or both of the \(\gamma_i\to\infty\). This is the way in which the mass terms accomplish the goal of giving us normalizable modes in the massless limit.

Now the canonical quantization is done by imposing the same commutation relations between the planar oscillators as we did in eq. \ref{eq:commalpha} for the transverse ones. The Hamiltonian in terms of the oscillators is again, after normal ordering of the oscillators,
\be \frac1\omega H_p = \sum_{n=1}^\infty (\alpha_{-n}\alpha_n + \tilde\alpha_{-n}\tilde\alpha_n) - (A + \tilde A) \ee
The intercept is given by
\be a_p= -\frac{1}{\omega}\langle H_p \rangle = A + \tilde A = -\frac12\sum_{n=1}^\infty \omega_n -\frac12\sum_{n=1}^\infty \tilde \omega_n \ee
The sum is calculated in the same method, by the contour integral, using the functions defined by eqs. \ref{eq:fpco} and \ref{eq:fpce} to sum over the eigenfrequencies. The form of the divergences in this sum is the same as for the transverse modes, and the renormalization procedure is the same. The result for both modes is as in \cite{Sonnenschein:2018aqf}, which is
\be \tilde A = -\frac{1}{2\pi}\int_0^\infty dy \log\left(1-e^{-2\delta y}\frac{(y-\gamma_0\beta_0)(y-\gamma_\ell\beta_\ell)}{(y+\gamma_0\beta_0)(y+\gamma_\ell\beta_\ell)}\right) \ee
\be A = -\frac{1}{2\pi}\int_0^\infty dy \log\left[1-e^{-2\delta y}\left(\frac{y^2-2\gamma_0\beta_0 y +\gamma_0^2(1+\beta_0^2)}{y^2+2\gamma_0\beta_0 y +\gamma_0^2(1+\beta_0^2)}\right)
\left(\frac{y^2-2\gamma_\ell\beta_\ell y +\gamma_\ell^2(1+\beta_\ell^2)}{y^2+2\gamma_\ell\beta_\ell y +\gamma_\ell^2(1+\beta_\ell^2)}\right)\right] \label{eq:acpei} \ee

One can get the leading order corrections by performing Zeta function regularization of the approximate solutions to eqs. \ref{eq:fpco} and \ref{eq:fpce}, which are
\be \tilde\omega_n = n + \frac{1}{3\pi}(n^3-n)(\frac{1}{\gamma_0^3}+\frac{1}{\gamma_\ell^3}) + \mathcal O(\gamma^{-5}) \ee
\be \omega_n = n - \frac{1}{6\pi}(n^3-n)(\frac{1}{\gamma_0^3}+\frac{1}{\gamma_\ell^3}) + \mathcal O(\gamma^{-5}) \ee
and find the result for the intercept using the Zeta function matches the expansion of the two integrals (see appendix \ref{app:Zeta} for the expressions). The result is that
\be a_p = A + \tilde A = \frac{1}{12}\left(1 - \frac{11}{120\pi}(\frac{1}{\gamma_0^3}+\frac{1}{\gamma_\ell^3}) \right) \label{eq:apce} \ee

\section{Quantizing the folded open string} \label{sec:quanto}
In this section we discuss the fluctuations around the rotating solution for the string with endpoint charges in a constant background magnetic field, introduced in section \ref{sec:rot_open_mag}. In this case there is a single fold at some point along the string. We place a mass there, analyze the spectrum of fluctuations, and compute the intercept as a function of the mass and the magnetic field. Finally, we can use the massless limit to obtain the result for the ordinary massless string.

\subsection{Classical solution with folding point mass}
We can write a rotating solution that has a mass at the point where the string is folded, in the following way. If the fold is at the point \(\sigma = \sigf\) and we insert the usual mass term into the action at that point, then in addition to the boundary conditions at the endpoints of the string,
\be T X^{\prime\mu} + q F^{\mu}{}_\nu \dot X^\nu = 0 \qquad \sigma = 0,\ell \ee
we have the condition at the folding point
\be T (X^{\prime\mu}|_{\sigma=\sigf^-}-X^{\prime\mu}|_{\sigma=\sigf^+}) + m \pa_\tau\left(\frac{\dot X^\mu}{\sqrt{-\dot X^2}}\right) = 0 \ee
There is a discontinuity in \(X^\prime\) but not in \(X\) or \(\dot X\). We write the usual rotating solution
\be X^0 = \tau \qquad X^1 = R(\sigma)\cos(\omega\tau) \qquad X^2 = R(\sigma)\sin(\omega\tau) \label{eq:sol_open_mass1}\ee
with \(R(\sigma)\) now defined piecewise as
\be R(\sigma) = \begin{cases} \frac1\omega\cos(\omega\sigma+\phi ) & \sigma \leq \sigf \\
															\frac1\omega\cos(\omega\sigma+\tilde\phi ) & \sigma \geq \sigf \label{eq:sol_open_mass2}\end{cases} \ee
We construct the solution such that it has a finite velocity at \(\sigma = \sigf < \ell\) and reduces to the normal classical solution of section \ref{sec:rot_open_mag} when the mass at the fold is taken to zero.

The boundary condition at \(\sigma=0\) determines \(\phi  = \arctan(\frac{qB}{T})\) as for the solution without the massive particle. The boundary condition at \(\sigma = \ell\) is that
\be \tan(\tilde\phi  + \omega\ell) = qB = \tan(\phi ) \ee
and the solution we take in order to get the correct result in the massless limit is
\be \omega\ell = \pi + \phi -\tilde\phi  \label{eq:wlo}\ee
The endpoint charges still move at the same finite velocity \(\beta_q = |\cos(\phi)| = |\cos(\omega\ell+\tilde\phi)|\).

The first condition on the solution at the folding point is that \(R(\sigma)\) is continuous there, namely that
\be \cos(\omega \sigf +\phi ) = \cos(\omega \sigf+\tilde\phi ) \ee
We cannot pick \(\phi  = \tilde\phi \) so we take the second possible solution
\be \tilde\phi  = 2\pi -\phi  - 2\omega \sigf \ee
with the \(2\pi\) added so that in the massless case the solution reduces to \(\phi  = \tilde\phi \) exactly (to be consistent with eq. \ref{eq:wlo} when \(\omega\ell=\pi\)).

With the mass there is a finite velocity of the folding point
\be \beta_f = -\cos(\omega \sigf+\phi ) \qquad \gamma_f = \frac{1}{\sin(\omega \sigf+\phi )} \ee
which is determined by the condition at \(\sigma=\sigma_f\),
\be \frac{2 T}{\gamma_f} = m\omega \gamma_f \beta_f \ee
This is the same equation as \ref{eq:bdc}, which is the force equation on the massive particle. It is useful to define the parameter
\be \delta \equiv \frac12\omega\ell = \omega \sigf + \phi  - \frac\pi2 \ee
in terms of which \(\beta_f = \sin\delta\), and which goes to \(\pi/2\) in the massless limit when \(\beta_f = 1\).

One condition that we should impose on the solution with the mass is that \(\sigf < \ell\), otherwise the mass is outside the domain where \(\sigma\) is defined. This condition is equivalent to \(\delta+\phi  > \frac\pi2\), which also means \(\beta_f > \beta_q\) - the folding point moves faster than the endpoints of the string. For a given value of the magnetic field, there is a value of the folding point mass above which we cannot find a solution that will obey the above condition.

The energy, angular momentum, and length of the open string with the massive particle are
\be E = \gamma_f m + \frac{\arcsin\beta_f}{\beta_f}TL \ee
\be J = \frac12\gamma_f m \beta_f L+\frac14 TL^2\frac{\gamma_f\arcsin\beta_f-\beta_f}{\gamma_f\beta_f^2}\ee
\be L = \frac{2\beta_f}{\omega} \ee
The expressions are the same as we find for the string with massive endpoints, except that now there is only one massive particle contributing to the energy and angular momentum. When the mass is small we can write the classical mass corrected Regge trajectory as
\be J_{cl}(E) = \alp E^2 \left[1 - \frac{8\sqrt\pi}{3}\left(\frac{m}{2E}\right)^{3/2} + \ldots \right] \ee

Again the goal is to compute the quantum correction to the above relation, which is the intercept
\be a = \langle J - J_{cl}(E) \rangle = -\frac{1}{\omega}\langle H_{ws}\rangle \ee
where \(H_{ws}\) is the worldsheet Hamiltonian for the fluctuations.

\subsection{Fluctuations}
\subsubsection{Transverse modes}

The \(D-3\) modes transverse to the plane of rotation are not affected directly by the rotation, nor the magnetic field, as those do not enter the equations of motion for the transverse fluctuations. They are only affected by the mass term we place to regularize the system. What we see is that when a finite mass is present at the fold, the spectrum of the transverse fluctuations becomes dependent also on the magnetic field, because of the modified boundary conditions. In the massless limit this dependence is removed.

The action for each of the \(D-3\) transverse fluctuation modes is
\be S_t = T\int d^2\sigma (\frac12\dot f_t^2 - \frac12 f_t^{\prime2}) + \gamma_f m \int d\tau \frac12 \dot f_t^2|_{\sigma=\sigf} \ee

The bulk equation of motion is the wave equation. We write the mode expansion
\be f_t(\tau,\sigma) = \alpha_0 + i \sqrt{\frac\alp2}\sum_{n\neq0}\frac{\alpha_n}{\omega_n}e^{-i\omega\omega_n\tau} f_n(\sigma) \label{eq:modexpo} \ee
For the eigenfunctions \(f_n\), which are just a combination of sine and cosine in this case, there are the two boundary conditions at the endpoints
\be f_n^\prime = 0 \qquad x = 0,\omega\ell \ee
as well as the conditions on the fold, which are continuity of \(f_n(\sigma)\) at \(\sigma_f\), as well as the discontinuity of the derivative given by
\be T(f_n^\prime|_{\sigf^+} - f_n^\prime|_{\sigf^-}) + \omega_n^2 \gamma_f m \omega f_n = 0, \qquad x=\omega \sigf \label{eq:bdcot} \ee
From the boundary conditions at the endpoints and the continuity requirement at the fold we get that the solutions must have the form
\be f_n(x) = \begin{cases} c_1 \cos(\omega_n x) & x \leq \omega \sigf \\ c_1 \frac{\cos((\frac\pi2+\delta-\phi)\omega_n)}{\cos((\frac\pi2-\delta-\phi)\omega_n)}\cos(\omega_n(x-2\delta))  & x \geq \omega \sigf \end{cases} \ee
where \(\phi  = \arctan(\frac{qB}{T})\) and \(\delta = \arcsin\beta_f\).

Then the last boundary condition at the fold \ref{eq:bdcot} holds provided that \(\omega_n\) satisfies
\be \sin\delta\sin(2\omega_n \delta) + \omega_n \cos\delta \left[\cos(2\omega_n \delta)+\cos(2(\frac\pi2-\phi)\omega_n)\right] = 0\ee
which we write also as
\be \beta_f\sin(2\omega_n \delta) + \frac{1}{\gamma_f}\omega_n \left[\cos(2\omega_n \delta)+\cos(2(\frac\pi2-\phi)\omega_n)\right] = 0 \label{eq:wto}\ee
This is in fact a function of the two relevant velocities in the problem: \(\beta_f\) of the fold, and \(\beta_q\) of the endpoint charges, since \(\delta = \arcsin\beta_f\) and \(\frac\pi2-\phi = \arcsin\beta_q\).

At the massless limit of \(\delta=\frac\pi2\) the equation reduces to
\be \sin(\pi \omega_n) = 0 \ee
or \(\omega_n = n\) with no dependence on \(\phi\), as expected. With masses, there is not only the dependence on the mass, but we have also introduced a dependence on the magnetic field.

In agreement with equation \ref{eq:ortho}, the eigenmodes satisfy
\be \omega^2(\omega_n^2-\omega_m^2) \int_0^{\ell} d\sigma f_m f_n = \left(f_n f_m^\prime-f_m f_n^\prime\right)|_0^\ell - \left(f_n \Delta f_m^\prime- f_m \Delta f_n^\prime\right)|_{x=\omega \sigf} \ee
Explicitly, after using the boundary conditions we are left with
\be  \int_0^{\ell} d\sigma f_m f_n + \frac{\gamma_f m}{T}f_m f_n|_{\sigma= \sigf} = \frac{2\pi}{\omega} (\delta_{m-n} + \delta_{m+n}) \label{eq:629}\ee

The worldsheet Hamiltonian derived for the transverse fluctuations is
\be H_t = T\int_0^\ell d\sigma (\frac12 \dot f_t^2 + \frac12 f_t^{\prime2}) + \frac12\gamma_f m \dot f_t^2|_{\sigma=\sigf} \ee
By inserting the mode expansion into the Hamiltonian and using eq. \ref{eq:629} we find that
\be H_t = \frac{\omega}{2} \sum_{n\neq0} \alpha_{-n}\alpha_n \ee
The quantization is done by imposing the commutator
\be [\alpha_m,\alpha_n] = \omega_n \delta_{m+n} \ee
And then, the intercept is the normal ordering constant
\be a_t = -\frac1\omega\langle H \rangle = -\frac12\sum_{n>0} {\omega_n} \ee
As for the closed string in the last section, we have two ways of regularizing this divergent sum. The first method is by use of the Zeta function. For this we need to write an approximate solution of the eigenfrequency equation \ref{eq:wto} for small masses. Then we have an expansion in \(\gamma_f^{-1}\), and the coefficient at each order can be calculated by Zeta function regularization. The details are in appendix \ref{app:Zeta_open}, where we also compare the result with the second method, which is converting the sum into a contour integral.

We use the contour integral over the same semicircular contour depicted in figure \ref{fig:contour}. The appropriate function to integrate over is the one defined by the equation for the eigenfrequencies we have in this case, eq. \ref{eq:wto}, so we define
\be g(z) = \sin\delta\sin(2\delta z) + z \cos\delta \left[\cos(2\delta z)+\cos((\pi-2\phi)z )\right] \ee

We compute the Casimir energy, regularized by introducing a cutoff \(\Lambda = \omega N\), 
\be E_C(\Lambda) = \frac\omega2 \sum_{n=1}^N \omega_n = \frac{\omega}{4\pi i}\oint dz z\frac{d}{dz}\log g(z) \ee
Evaluating the integral on the semicircle at large \(N\)
\be I_{sc}(\Lambda) = \frac{\omega N^2}{4\pi}\int_{-\frac\pi2}^{\frac\pi2} d\theta e^{2i\theta}\frac{g^\prime(Ne^{i\theta})}{g(Ne^{i\theta})} \to \frac{\omega \delta}{\pi}N^2 + \frac{\omega}{2\pi}N = \frac{1}{2\pi}\Lambda^2 \ell + \frac{1}{2\pi}\Lambda \ee
We find no contributions that remain finite at infinite \(\Lambda\). The integral on the imaginary axis we integrate by parts, and it is
\be I_{im}(\Lambda) = -\frac{\omega}{4\pi}\int_{-N}^N dy y \frac{d \log g(iy)}{dy} = -\frac{\omega}{4\pi} N \left(\log g(iN) + \log g(-iN)\right) + \frac{\omega}{4\pi}\int_{-N}^N dy \log g(iy) \ee
We first look at only the divergent terms when \(N\) is large. To find them we can take the asymptotic forms of \(g(iy)\) when \(|y|\gg1\),
\be g^+(y) = \frac{i}{2} e^{2\delta y}(y\cos\delta + \sin\delta)\,, \qquad g^-(y) = \frac{i}{2}e^{-2\delta y}(y\cos\delta-\sin\delta) \label{eq:gasym}\ee
where \(g^+\) is the asymptotic form at \(y\) large and positive, and \(g^-\) likewise at negative \(y\).  We have used the condition that there is a fold in the classical solution, \(\delta > \frac\pi2 - \phi\), and as a result there is no dependence on the magnetic field in the divergent terms.

If we evaluate the integral on this function,
\be I^{div}_{im}(\Lambda) = -\frac{\omega}{4\pi} N \left(\log g^+(N) + \log g^-(N)\right) + \frac{\omega}{4\pi}\int_{0}^N dy \log g^+(y) + \frac{\omega}{4\pi}\int_{-N}^0 dy \log g^-(y) \ee
then we get that the divergent part is
\be I^{div}_{im}(\Lambda) = -\frac{1}{4\pi}\Lambda^2\ell -\frac{1}{2\pi}\Lambda +\frac{\omega\tan\delta}{2\pi}\log\left(\frac{\Lambda}{\omega\tan\delta}\right) + \mathcal O(\frac{1}{\Lambda}) \ee
Combining the results for \(I_{sc}\) and \(I_{im}\), we find that
\be E_C(\Lambda) = \frac{1}{4\pi}\Lambda^2 \ell +\frac{T}{\pi\gamma_f m}\log\left(\frac{\gamma_f m\Lambda}{T}\right) + \ldots \ee
where the ellipses signify terms that do not depend on the cutoff as we take it to infinity.

The form of the divergences is the same as for the case of a rotating string with massive endpoints, as well as the closed string analyzed section \ref{sec:quantc}. Again the subtraction of the divergences can be done by redefining the string tension and the mass, or by subtracting the infinite string's contribution for the Casimir force.

This can be achieved before integration by subtracting the divergent parts written in their integral form. The semicircle integral contains only divergent terms and is subtracted, such that the intercept can be written as
\be a_t = -\frac{1}{\omega}E_C^{ren} = -\frac1{2\pi}\int_0^\infty dy \log \left(\frac{g(y)}{g^+(iy)}\right) \ee
where \(g^+\) is the asymptotic form of the function at large, positive \(y\).\footnote{The contribution from negative \(y\) is the same as the positive, so the intercept is here written as twice the integral over positive \(y\).} This formula is actually quite general and can be used for any of the cases analyzed in this paper.

At this point we see that if one takes the form of \(g^+(y)\) written in eq. \ref{eq:gasym} and inserts it in the integral then the answer one would get for \(a_t\) has a divergence as the magnetic field is taken to be small. We can see this by looking at the leading order correction at large \(\gamma_f\),
\be a_t = \frac{1}{24}\left(1 + \frac{1-\frac{3}{\sin^2\phi}}{\pi}\frac{1}{\gamma_f} + \ldots\right) \ee
which is clearly problematic at small \(\phi\). The phase is \(\phi = \arctan(qB/T) \approx qB/T\) at small fields, so the divergence is proportional to \(T/qB\). The divergence probably reflects a subtlety in the subtraction due to the fact that we renormalize \(T\). It turns out that the subtraction should be done using a form that keeps a subleading term that depends on \(\phi\), which is exponentially smaller than the leading term, and does not contribute to the divergent terms at finite \(\phi\). The form we use is
\be g^+ = \frac{i}{2}\left(e^{2\delta y}(y\cos\delta+\sin\delta) + e^{2(\frac\pi2-\phi)y}y\cos\delta\right) \ee
Only then one gets the correct result, which has no problems at small \(\phi\) and matches exactly with the Zeta function answer. The final result for the transverse intercept with finite masses is then
\be a_t = -\frac1{2\pi}\int_0^\infty dy \log\left(1-e^{-2(\delta-\phi+\frac\pi2)y}\frac{-y - e^{-2(\delta+\phi-\frac\pi2)y}(y-\gamma_f\beta_f)}{y+\gamma_f\beta_f +  e^{-2(\delta+\phi-\frac\pi2)y} y}\right) \label{eq:a_t_open_phi}\ee
At small mass,
\be a_t = \frac{1}{24}\left(1 + \frac{1+\frac{3}{\phi^2}-\frac{3}{\sin^2\phi}}{\pi}\frac{1}{\gamma_f} + \ldots\right) \ee
The leading order term now vanishes at the \(\phi=0\) limit, which is the expected result. Higher order terms and their derivation from the integral and Zeta function are in appendix \ref{app:Zeta_open}.

\subsubsection{Planar mode}
We now derive the action and equations of motion for the planar fluctuations around the solution with the massive fold. We define
\be f_p = \begin{cases} \frac{\cot(\omega\sigma+\phi )}{\omega}\delta\theta & 0\leq \sigma \leq \sigf \\ -\frac{\cot(\omega\sigma+\tilde\phi )}{\omega}\delta\theta & \sigf \leq \sigma \leq \ell \end{cases} \ee
The relative minus sign in the definitions is needed if we want the definition of \(f_p\) to be continuous at the folding point \(\sigma=\sigf\).

The full action for the system, including the massive particle on the folding point, is
\be S = S_{NG} + q \int d\tau  A_\mu \dot X^\mu|_{\sigma=0} - q \int d\tau  A_\mu \dot X^\mu|_{\sigma=\ell} - m \int d\tau \sqrt{-\dot X^2}|_{\sigma=\sigma_f} \ee

The bulk part of the action is obtained from expanding the Nambu-Goto action and is given in eq. \ref{eq:Sp}. On the boundary points we have
\be S_b(0) = \int d\tau \left[T\left(\frac{\omega}{\sin(2\phi  )}f_p^2+\frac12 \omega\cot\phi   f_r^2 + \cos\phi   f_r \dot f_p\right)+ q B\left(\frac12\omega f_r^2+\sin\phi   f_r \dot f_p\right)\right]\ee
The first set of terms come from the NG action in the bulk, the second set proportional to \(q B\) from the boundary interaction with the magnetic field. Using the condition \(q B = T\tan\phi  \), we can reduce the boundary action to the form
\be S_b(0) = \frac{T q B}{2}\int d\tau \left(\gamma_q^2\omega(f_p^2+f_r^2)+\gamma_q(f_r \dot f_p-f_p\dot f_r)\right) \ee
At the other endpoint of the string there is a charge \(-q\), and the action is different only by overall sign
\be S_b(\ell) = -\frac{T q B}{2}\int d\tau \left(\gamma_q^2\omega(f_p^2+f_r^2)+\gamma_q(f_r \dot f_p-f_p\dot f_r)\right) \ee
The endpoints move at a finite velocity given by
\be \gamma_q = \frac{1}{\sin\phi } = -\frac{1}{\sin(\omega\ell+\tilde\phi )} \ee
Lastly, there are the terms at the folding point
\be S_b(\sigf) = \gamma_f m \int d\tau \left(\frac12 \dot f_p^2 + \frac12\dot f_r^2+\frac12 \mu_p^2 f_p^2+\frac12\mu_r^2 f_r^2+c f_r \dot f_p\right)\ee
with
\be \gamma_f = \frac{1}{\sin(\omega \sigf+\phi )} = -\frac{1}{\sin(\omega \sigf+\tilde\phi )} \ee
and the parameters
\be \mu_p^2 = \gamma_f^2\omega^2\,, \qquad \mu_r^2 = (2\gamma_f^2-1)\omega^2 \,, \qquad c = 2\gamma_f \omega\ee

The variation of the action on the boundaries w.r.t. \(f_r\) results in the following equations
\be \gamma_q^2 \omega f_r + \gamma_q\dot f_p = 0, \qquad \sigma = 0, \ell \label{eq:658}\ee
\be -\ddot f_r + \mu_r^2 f_r + c \dot f_p = 0\,,\qquad \sigma = \sigf \ee 
The variation of \(f_p\) yields the bulk equation of motion as well as the boundary equations
\be T f_p^\prime + TqB(\gamma_q^2\omega f_p - \gamma_q \dot f_r) = 0, \qquad \sigma = 0, \ell \ee
and at the fold
\be T(f_p^\prime|_{\sigf^+} - f_p^\prime|_{\sigf^-}) + \gamma_f m (- \ddot f_p + \mu_p^2 f_p - c\dot f_r) = 0, \qquad \sigma = \sigf \ee

We write the Fourier expansion for \(f_p\) of the same form as in eq. \ref{eq:modexpo}. The mode \(f_r\), which lives only at the boundary points and the fold, can be expanded with the same set of eigenfrequencies
\be f_r =  i\sqrt{\frac\alp2}\sum_{n} \frac{\alpha_n}{\omega_n} f_r^{(n)} e^{-i\omega \omega_n\tau} \ee
where the coefficients \(f_r^{(n)}\) are just constants (defined separately at each boundary point), and the \(\alpha_n\) are the same that appear in the expansion for \(f_p\).

We can solve the equations for \(f_r\) at each of the boundary points to get a boundary condition only on \(f_p\). The solutions at \(\sigma = 0\) and \(\sigma = \ell\) are obtained immediately from equation \ref{eq:658}. At \(\sigma = \sigf\) we have
\be f_r^{(n)} = \frac{i\omega \omega_n c}{\omega^2\omega_n^2+\mu_r^2}f_n(\sigf) \ee
where \(f_n(\sigma)\) are the eigenmodes of the planar mode \(f_p\). 

We can write then the boundary conditions in terms of \(f_n\) only, which now we write as a function of \(x = \omega\sigma\):
\be f_n^\prime + q B(\gamma_q^2 - \omega_n^2)f_n = 0\,, \qquad x = 0,\omega \ell \ee
\be f_n^\prime|_{\sigma_f^+} - f_n^\prime|_{\sigma_f^-} + \frac{\gamma_f m}{T\omega}(\omega^2\omega_n^2+\mu_p^2 - \frac{c^2\omega^2\omega_n^2}{\omega^2\omega_n^2+\mu_r^2})f_n = 0\,, \qquad x=\omega \sigf\ee
In addition to the last condition we demand that \(f_n\) itself is continuous at \(x = \sigf\).

The general solution of the bulk equations of motion is
\be f_n(x) = \begin{cases} c_1 s_1(x+\phi ) + c_2 s_2(x+\phi ) \qquad 0\leq x \leq \omega \sigf \\
										  c_3 s_1(x+\tilde\phi ) + c_4 s_2(x+\tilde\phi ) \qquad \omega \sigf \leq x \leq \omega \ell\end{cases} \ee
where \(s_1\) and \(s_2\) are the independent solutions of the bulk equation of motion, given in eqs. \ref{eq:s1}-\ref{eq:s2}. The explicit form of the solution that satisfies the boundary conditions at \(\sigma=0,\ell\) is
\be f_n(x) = \begin{cases} \tilde c_1\left(\cos(\omega_n x)\cot(x+\phi )+\omega_n\sin(\omega_n x)\right) & 0\leq x \leq \omega \sigf \\
										\tilde c_3\left(\cos[\omega_n (x-2\delta)]\cot(x+\tilde\phi )+\omega_n\sin[\omega_n (x-2\delta)]\right) & \omega \sigf\leq x \leq \omega \ell\end{cases} \ee
Then \(\tilde c_3\) can be related immediately to \(\tilde c_1\) by the continuity requirement at \(x=\omega \sigf\). The remaining equation is satisfied provided that \(\omega_n\) is
\begin{multline}
0 = 2 \omega_n \left(1-\omega_n^2\right) \cos^3\delta \cos (\omega_n (\pi -2 \phi ))+\\+\omega_n \cos\delta \left(3 \cos (2 \delta )+2 \omega_n^2 \cos ^2\delta-5\right) \cos (2 \omega_n \delta)+\\
+\sin\delta  \left(\cos (2 \delta )+6 \omega_n^2 \cos ^2\delta-3\right)\sin (2 \omega_n \delta)
\end{multline}
In terms of \(\beta_f = \sin\delta\) and \(\gamma_f = \sec\delta\),
\begin{multline}
\frac{1}{\gamma_f^3}\left(\omega_n^3+(3-4\gamma_f^2)\omega_n\right)\cos(2\omega_n \delta) +\left(\frac{3\beta_f}{\gamma_f^2}\omega_n^2-\beta_f(1+\beta_f^2)\right)\sin(2\omega_n \delta) + \\
-\frac{1}{\gamma_f^3}(\omega_n^3-\omega_n)\cos (2(\frac\pi2 -\phi)\omega_n) = 0 \label{eq:wpo}
\end{multline}

When the mass at the fold is small we write can write an approximate solution when \(\frac{1}{\gamma_f}\ll 1\),
\be \omega_n = n + \frac{1-3\cos(2n\phi)}{6\pi}(n^3-n)\frac{1}{\gamma_f^3} + \mathcal O(\frac{1}{\gamma_f^5}) \label{eq:omega_p_approx}\ee


Using eq. \ref{eq:ortho} with the appropriate boundary terms for the present case, we can define the inner product of two eigenfunctions as
\begin{multline}
\int_0^{\omega\ell} dx f_m f_n + \tan\phi \left(f_m f_n|_{x=0}-f_m f_n|_{x=\omega\ell}\right) +\\+ \frac{\gamma_f m \omega}{T}\left(1-\frac{c^2\mu_r^2}{(\omega^2\omega_m^2+\mu_r^2)(\omega^2\omega_n^2+\mu_r^2)}\right)f_m f_n|_{x=\omega \sigf} = \frac{\pi}{\omega} (\delta_{m-n} + \delta_{m+n}) 
\end{multline}
As with the planar mode on the folded closed string, the extra term at the folding point lets us take the massless limit smoothly, even as the eigenfunctions diverge at the point. We see the same canceling out of divergences between the bulk integral and the added terms at the folding point.


The computation of the intercept follows the one outlined in the previous subsection for the transverse modes. Following the contour integral method, the result is
\be a_p = -\frac{1}{2\pi}\int_0^\infty dy \log\left(1-e^{-2(\delta-\phi+\frac\pi2)y}
\frac{(y^3+y)- e^{-2(\delta+\phi-\frac\pi2)y}[y^3-3\gamma_f\beta_f y^2+(4\gamma_f^2-3)y-\gamma_f^3\beta_f(1+\beta_f^2)]}
{y^3+3\gamma_f\beta_f y^2+(4\gamma_f^2-3)y+\gamma_f^3\beta_f(1+\beta_f^2)-e^{-2(\delta+\phi-\frac\pi2)y}(y^3+y)}\right) \label{eq:a_p_open}\ee
Where again we need to subtract a \(\phi\)-dependent term to get the correct result.

The Zeta function regularization of the sum over \(\omega_n\), using the approximate solution of eq. \ref{eq:omega_p_approx}, we can get the result
\be a_p = \frac{1}{24}-\frac{1}{24 \pi }\left(\frac{9}{4 \phi ^4}+\frac{3}{2 \phi ^2}-\frac{9}{4\sin^4\phi} +\frac{11}{60}\right)\frac{1}{\gamma_f^3} + \ldots \label{eq:apo} \ee
The derivation of this and the matching integral form are in appendix \ref{app:Zeta_open}.

Interestingly, the \(\phi\to0\) limit now reproduces the result from the string with massive endpoints, \(a_p = \frac{1}{24}+\frac{11}{720\pi}\gamma_f^{-3}\), even though this limit takes us out of the assumed range of the parameters.
	

\section{The intercept of non-critical rotating strings} \label{sec:PS}

In dimensions other than the critical dimension of \(D = 26\) there is another contribution to the intercept of the string from the Polchinski-Strominger (PS) term in the effective string action \cite{Polchinski:1991ax}. In the orthogonal gauge, and in terms of the coordinates \(\sigma_\pm = \tau\pm\sigma\), this added term is
\be S_{PS} = \int d\tau\mathcal L_{PS} = \frac{26-D}{24\pi}\int d^2\sigma \frac{(\pa_+^2 X\cdot \pa_- X)(\pa_-^2 X\cdot \pa_+ X)}{(\pa_+ X\cdot \pa_- X)^2} \ee
This is understood as the leading correction to the Nambu-Goto action in the effective string theory \cite{Aharony:2010cx,Aharony:2013ipa,Dubovsky:2012sh,Hellerman:2014cba}, when expanding around a classical solution with a length parameter \(L\). The action is expanded then in powers of \(\ell_s/L\), where \(\ell_s = \sqrt{\alp}\) is the intrinsic length scale of the string theory. The coefficient of the PS term is fixed such that the conformal symmetry in dimensions other than 26 is preserved, up to higher order terms in the long string expansion.

The effect of the PS term is to add a correction to the Hamiltonian whose effect at leading order we can compute simply by evaluating the PS Lagrangian on the classical solution,
\be E_{PS} =  -\int d\sigma \mathcal L_{PS}(\bar X) \ee
Depending on the classical solution, we may find that this diverges and needs renormalization, and this is indeed the case for rotating string solutions.

In the original paper of Polchinski and Strominger \cite{Polchinski:1991ax}, the PS term was derived by starting from the Liouville action
\be S_L = \frac{D-26}{24\pi}\int d^2\sigma \pa_+ \varphi \pa_- \varphi\ee
This term originates from the Weyl anomaly in the Polyakov formulation of the bosonic string theory, where we have an independent worldsheet metric \(\gamma_{ab}\) and the classical Weyl symmetry \(\gamma_{ab} \to e^{\Omega(\tau,\sigma)}\gamma_{ab}\). In \cite{Polyakov:1981rd} the Liouville action appears as a change of the path integral measure \(\mathcal D \gamma\) under Weyl transformations, when fixing the conformal gauge \(\gamma_{ab} = e^{\varphi}\eta_{ab}\). In \cite{Polchinski:1991ax} it was proposed to start from the Nambu-Goto action and identify \(h_{ab} = e^{\varphi}\eta_{ab}\), where \(h_{ab}\) is now the induced metric on the worldsheet (assumed to be in the orthogonal gauge). Substituting \(\log h_{+-}\) for \(\varphi\) in the Liouville action gives precisely the PS term. The idea of \cite{Polchinski:1991ax} is to write an action only using the fluctuations of the string, but there is a more geometric interpretation of this term, from which we see that we could encounter problems at points where the determinant of \(h_{ab}\) vanishes, or where the worldsheet curvature diverges.
%
 
For a rotating string we find that the PS term adds a contribution to the intercept which is
\be a_{PS} = -\frac{1}{\omega} E_{PS} = \frac{D-26}{24\pi} \omega\int d\sigma \cot^2(\omega\sigma+\phi) \ee
with the appropriate angular velocity \(\omega\), phase \(\phi\), and integration boundaries depending on the solution. The integrand diverges whenever the argument of the cotangent is a multiple of \(\pi\), which are the points in the solution moving at the speed of light, and the points at which the worldsheet curvature diverges.

In \cite{Hellerman:2013kba} the intercept of the rotating Neumann open string was computed with appropriate boundary terms used to cancel out the divergence. In \cite{Sonnenschein:2018aqf} we quantized the open string with massive endpoints and used the mass terms to regularize the system. 
Adding masses to the folding points, the renormalization procedure for rotating folded strings will be identical as for the open string.

\subsection{Closed folded string}
For the folded rotating solution of the closed string with masses at the folding points, presented in section \ref{sec:closed_masses}, the PS correction is given by
\be E_{PS} = \frac{26-D}{24\pi}\omega\left(\int_{-\delta}^0 dx \cot^2(x-\phi)+\int_0^\delta dx \cot^2(x+\phi)\right) \ee
which is
\be E_{PS} = \frac{26-D}{12\pi}\omega\left(\cot\phi-\cot(\delta+\phi)-\delta\right) = \frac{26-D}{12\pi}\left(\frac{2T}{\gamma_0 m_0}+\frac{2T}{\gamma_\ell m_\ell}-\frac{\delta^2}{\ell}\right) \ee
The result is essentially the same as for a string with massive endpoints analyzed in \cite{Sonnenschein:2018aqf}, with the tension effectively doubled, \(T\to 2T\).

In \cite{Sonnenschein:2018aqf} it was argued that the terms which diverge in the massless limit (\(\gamma m\to 0\)) must be subtracted from the final result. As with the contribution from the fluctuations, the second term that goes like \(1/\ell\) gives the PS intercept. The subtraction can be done by renormalization of the masses on the folds. The result is then that for any finite endpoint mass,
\be a_{PS} = -\frac{1}{\omega}E_{PS}^{(ren)} = \frac{26-D}{12\pi}\delta = \frac{26-D}{12\pi}\left(\arcsin\beta_0 + \arcsin\beta_\ell\right) \label{eq:aPSclosed} \ee
In the massless limit, \(a_{PS} = (26-D)/12\), so adding all the contributions from the fluctuations and the PS term, the full intercept of the closed string at any \(D\) becomes
\be a = \frac{D-2}{12} + \frac{26-D}{12} = 2 \ee
as it is in the critical dimension. This is an extension of the result \(a=1\), independently of the dimension, that was found in \cite{Hellerman:2013kba,Sonnenschein:2018aqf} for the open string.

However, if finite mass terms are generated at the folds then the intercept is a function of those masses. In that case, the PS contribution can be the dominant correction coming from the masses, especially far from the critical dimension. 

The full intercept of the closed string, including \(a_{PS}\) and the contribution of all the fluctuation modes, given in eqs. \ref{eq:atce} and \ref{eq:apce}, is, in the small masses expansion
\begin{align} \label{eq:aclosed} a &= (D-3)a_t + a_p + a_{PS} = \\ &= 2 + \frac{3D-55}{24\pi}(\epsilon_1+\epsilon_2)+\frac{D-3}{24\pi^2}(\epsilon_1+\epsilon_2)^2+\frac{D-3}{24\pi^3}(\epsilon_1+\epsilon_2)^3+\frac{8D-495}{1440\pi}(\epsilon_1^3+\epsilon_2^3) + \ldots \nonumber
 \end{align}
where \(\epsilon_1 = \frac{1}{\gamma_0}\) and \(\epsilon_2 = \frac{1}{\gamma_\ell}\).

It is interesting that the leading order correction is negative for any \(D\leq18\), but it is positive when \(D>18\), including in the critical dimension. 
For ordinary bosonic strings, there is a condition that requires \(a\leq 2\), otherwise there are negative norm states in the spectrum. With the inclusion of masses, one should verify that we do not violate any consistency condition on the theory when \(D>18\) and we get an intercept greater than 2. As a semiclassical description of long strings there is no apparent problem with a positive sign correction, but one should check that there is no violation of some no ghost theorem on the worldsheet when we added the mass terms.

\subsubsection{Closed string in two planes of rotation} \label{sec:PSc2}
The calculation of the PS term on the rotating closed string in two planes of rotation (presented here in section \ref{sec:classicalTwoPlanes}) was already done in \cite{Hellerman:2013kba}. We repeat it here.

Inserting the solution of eq. \ref{eq:rotsol2b} into the PS term, the intercept will be given by
\be a_{PS} = \frac1\omega\int_{-\ell}^\ell d\sigma \mathcal L_{PS} = \frac{D-26}{24\pi} \int_{-\pi}^\pi dx \left(\frac{\cos(2\xi)\sin(2x)}{1-\cos(2\xi)\cos(2x)}\right)^2 \label{eq:711} \ee
For any finite ratio of \(J_1/J_2\) the string has no folds, no points moving at the speed of light, the denominator of \(\mathcal L_{PS}\) is always finite, and so is the result for the intercept
\be a_{PS} = \frac{D-26}{12}\left(\frac{1}{|\sin(2\xi)|}-1\right) = \frac{D-26}{24}\left(\sqrt{\frac{J_1}{J_2}}+\sqrt{\frac{J_2}{J_1}}-2\right)\ee
which is what was found in \cite{Hellerman:2013kba}. 

This is finite for any finite \(J_1\) and \(J_2\) but is very large when one of the angular momenta is much smaller than the other, which is a worrying result. Taking one of the angular momenta, say \(J_2\), to zero gives a divergent result, but below that divergence we do have the constant part of \((26-D)/12\) which we found for the folded string, starting from \(J_2 = 0\) and renormalizing the divergence with masses.

It is interesting to consider this solution in the case where \(J_1/J_2 \gg 1\). By increasing this ratio we can get arbitrarily close to having a folding point that moves at the speed of light. One conjecture of what happens then is that effective mass terms are again generated near that extremal point, in such a way that will allow us to renormalize the divergent part and get a result that has a finite limit in the \(J_2\to 0\). Specifically we could expect a result that goes to \(a_{PS} = (26-D)/12\) when \(J_2\to0\), and then the form of eq. \ref{eq:711} suggests the leading order correction being of order \(\sqrt{J_2/J_1}\). On the other hand, as shown in eq. \ref{eq:curvTwoPlanes}, the worldsheet curvature evaluated on the classical solution, does not have a smooth limit at the extremal points, being large and negative for finite \(J_1/J_2\) but positive infinity in the limit \(J_2\to0\). This behavior might be modified by adding mass terms.

\subsection{Open string}
For the open string, we evaluate the PS Lagrangian on the solution of a rotating open string in a magnetic field with a mass on the folding point, defined in eqs. \ref{eq:sol_open_mass1}-\ref{eq:sol_open_mass2}. The correction to the energy is
\be E_{PS} = \frac{26-D}{24\pi}\omega^2\left(\int_0^{\sigma_f} d\sigma \cot^2(\omega\sigma+\phi ) + \int_{\sigma_f}^\ell d\sigma \cot^2(\omega\sigma+\tilde\phi )\right) \ee
After integration, this is the same form as for the folded closed string, with the difference that now there is only one folding point and one mass term instead of two.
\be E_{PS} = \frac{26-D}{12\pi} \omega (\tan\delta-\delta) = \frac{26-D}{12\pi}\left(\frac{4 T}{\gamma m}-\frac{2(\arcsin\beta_f)^2}{\ell}\right)\ee
The term that diverges as \(\gamma m\to 0\) has to be subtracted as before, with the renormalized result for the PS intercept being
\be a_{PS} = -\frac1\omega E_{PS}^{(ren)} = \frac{26-D}{12\pi}\delta = \frac{26-D}{12\pi}\arcsin\beta_f \label{eq:aPSopen}\ee
The result depends only on the velocity of the folding point, so there is no dependence on the magnetic field. In the massless limit \(a_{PS} = \frac{26-D}{24}\), such that, for the open string with no masses rotating in a magnetic field, the full intercept is given by
\be a = \frac{D-2}{24} + \frac{26-D}{24} = 1 \ee
for any \(D\) and any value of the magnetic field \(B\).

In the critical dimension, as long as the two charges on the endpoints of the string sum up to zero, the intercept remains \(a=1\) independent of the external field \cite{Abouelsaood:1986gd}. The result a rotating string with no endpoint charges in \(D\) dimensions is also \(a=1\) as shown in \cite{Hellerman:2013kba,Sonnenschein:2018aqf}. Our result shows that combining the two by having the non-critical string in a magnetic field does not change the intercept.

Since a finite mass may develop at the folding point, we can write the full result for finite masses, adding the results of eqs. \ref{eq:a_t_open_phi}, \ref{eq:a_p_open}, and \ref{eq:aPSopen}. The first order correction is
\be a = (D-3)a_t + a_p + a_{PS} = 1 + \frac{3D-55+3(D-3)\left(\frac{1}{\phi^2}-\frac{1}{\sin^2\phi}\right)}{24\pi}\frac{1}{\gamma_f} + \ldots \label{eq:aOpen} \ee
Now the modification of the masses also adds a dependence on the magnetic field in the intercept, even when the total charge of the string is zero. It is not immediately apparent from the expression, but for any \(D \leq 26\) and any \(\phi\) the leading order correction is negative.

We compute the corrections up to order \(\gamma_f^{-3}\), but since the coefficients are quite complicated functions of \(\phi\) we write them separately in appendix \ref{app:Zeta_open}.

 
\section{Summary and open questions} \label{sec:summary}
In this paper we analyzed folded bosonic rotating string configurations, focusing on the rotating solutions corresponding to states on the leading Regge trajectories of closed strings, as well as open string in a magnetic field, both of which exhibit folds.

We determined the spectrum of fluctuations around these strings, and calculated the quantum correction to the Regge trajectory, which is the intercept. To address the divergences that arise in this semiclassical quantization of strings with folds, we insert mass terms at the folding points. Starting from the action of a folded string with folding point masses, we write the modified classical solution in the presence of the masses, and perform the canonical quantization of the fluctuations around this new configuration. The problem of divergences at the folds, elaborated in section \ref{sec:fluctuations}, is solved as the masses slow down the folding point from the speed of light, making everything finite at the folding points. The system also has a well defined limit when the masses go to zero where the would be divergences are canceled out by the contribution from the masses.

The problem of divergences in the quadratic action for the fluctuations is quite subtle, leading to a problem in defining the normalization of the eigenmodes. In non-critical theories there is a more obvious divergence in the Polchinski-Strominger term that is added to effective string action. 

In section \ref{sec:quantc} we quantized the fluctuations around a rotating folded closed string in \(D\) dimensions. In order to do that, we modified the solution by starting from a system with massive particles placed at the two folding points of the string, and quantized the fluctuations in the presence of these masses. In this way we regularize the divergences associated with the folding point, and then there is a smooth massless limit which gives us the result for an folded closed string with no masses.

In generic theories, one could expect that such mass terms will emerge in the effective long string description, and it that case the masses would lead to finite corrections to the asymptotic Regge trajectory and the intercept which we calculated in this work. The result for the intercept closed string in \(D\) dimensions, including the correction from the PS term in the non-critical theory is in eq. \ref{eq:aclosed}.

In section \ref{sec:quanto} we performed a similar analysis for an open string with endpoint charges \(+q\) and \(-q\) rotating in a magnetic field. This solution has a fold, unless there are endpoint masses of the string in addition to charges. In this paper we analyzed the solution with a mass only at the folding point, and found the corrections related to the mass. The mass also introduces a novel dependence on the magnetic field through the modified boundary conditions, even though the total charge of the string is zero. The result for the intercept is in eq. \ref{eq:aOpen}.

For both open and closed strings, the result in the massless limit is that the intercept is independent of the dimension, a result which generalizes what was found in \cite{Hellerman:2013kba} to the closed string, and to an open string with opposite charges rotating in a magnetic field.

In addition to the divergences associated with the folding point, there are also the more ordinary UV divergences of the worldsheet theory, that in this paper appear in computing the intercept as the normal ordering constant in the Hamiltonian, the sum of eigenfrequencies. We show in sections \ref{sec:quantc} and \ref{sec:quanto} how the method used in \cite{Lambiase:1995st,Sonnenschein:2018aqf}, of converting the sum into a contour integral and renormalizing it works for the cases analyzed in this paper. The result are equivalent to using order by order Zeta function regularization, as detailed in appendix \ref{app:Zeta}. The renormalization can be thought of as either subtracting the Casimir force of an infinitely long string, or as a redefinition of the tension and masses.

While we discuss the massless limit of our result often, we cannot rule out the appearance of finite corrections associated with the masses in generic theories that have an effective string description. We think not only of quantizing the strings of string theory in a semiclassical way, but of a larger class of theories that have strings in some limit. This includes Yang-Mills theory and QCD. The description of effective long strings is constrained and the folding point masses are then generic objects that might appear in that description.

While endpoint masses of the string seem very natural to include in QCD, being associated with masses of quarks, one can think now also of folding point masses that will affect the closed string states, the glueballs. In a generic effective string theory, these masses will be a function of the underlying theory. It will be interesting to see if there is evidence of mass corrections in YM by comparing with results from the lattice.
 
The topic of folded strings has not received much attention in the space of questions associated with string theories. Nevertheless, it is quite plausible that folded strings are important objects in various domains of physics from hadrons and cosmic strings and all the way to black holes. Thus, there is a large unexplored region of questions associated with folded strings. Some arise from this current work, while others are more general. Let us give a partial list of open questions.
\begin{itemize}
\item In holography, endpoint masses of the string arise from ``vertical'' segments of the strings, which are along the holographic radial dimension \cite{PandoZayas:2003yb}. The folded rotating closed string solutions in confining backgrounds, as we saw in section \ref{sec:classicalConf}, are located at the holographic wall, at some constant \(r=r_0\). It will be interesting to see if the folding point masses on the closed string can correspond to some deformation of the holographic solution, i.e. by taking non-trivial solutions of eqs. \ref{eq:361}-\ref{eq:362}.

\item As mentioned in section \ref{sec:classical}, there exist rotating solutions with angular velocity \(\omega = n\frac\pi\ell\) for any integer \(n\), and by taking \(n>1\) we add more folds to the string. In this paper we did not analyze the semiclassical quantization of these multiply folded strings, choosing to focus on \(n=1\) which describes the states on the leading Regge trajectories. On the other hand, if one wants to consider all states in the string spectrum, one has to consider also the multiply folded solutions.

\item In appendix \ref{app:Zeta} we see some interesting formulas relating sums of Zeta functions to integrals. One question that arises from that is whether one can construct strings with special boundary conditions and in this way generate more identities of that sort. For instance, in \ref{app:Zeta_open} we deform the boundary conditions of the string in a magnetic field by adding a small mass, so we can write an expansion for the spectrum of eigenfrequencies the form \[\omega_n = n + c_1(n;\phi) \epsilon + c_2(n;\phi)\epsilon^2+\ldots\]
In that case \(\phi\) is related to the magnetic field and \(\epsilon\) to the small mass at the folding point, and the coefficients are fixed by the boundary condition. One can try and engineer boundary conditions for the string that will result in coefficient functions that will lead to interesting identities when computing the intercept as the renormalized sum over \(\omega_n\) in both methods.

\item
It would be interesting to estimate the weight of folded strings in the total space of string configurations. String theories are characterized by exponential growth of the density  of states. Does this apply also for folded strings?  How does the weight of folded strings depend on the parameters of a particular theory, such as the string coupling, the topology and geometry of the target space, etc.? Is that dependence the same for open strings and closed strings?

\item
In this paper we discuss bosonic strings. One may wonder if one can also supersymmetrize the notion of a fold, namely write the superstring equivalent of the condition $|det(J^\mu_a)|= 0$.

\item
In this paper we have concentrated on the classical and quantum rotating folded strings in flat spacetimes. Folded closed strings have been analyzed intensively in $Ads_5\times S^5$ both in the classical as well as one loop level \cite{Giombi:2010bj}. It will be interesting to apply the method developed here to regulate the induced metric scalar curvature  also for case of rotating stringy configurations in curved backgrounds like $Ads_5\times S^5$ and the large variety of other holographic backgrounds, in particular confining ones.

\item
In section \ref{sec:classicalMaldacena} we have presented an example of non-critical folded strings which are non-rotating. In \cite{Attali:2018goq} it was argued that  these type of strings introduced in \cite{Maldacena:2005hi} may play an important role in the description of the interior of black holes. In these papers only the classical solutions were considered. It will be very interesting to address the quantization of those strings and in particular whether the method of renormalizing the theory using massive particles on the folds can be applied also to these non-rotating folds.


\item
In \cite{Sonnenschein:2017ylo} the decay processes of open and closed hadronic strings have been analyzed and the corresponding total and partial decay widths have been computed. The analysis in that paper was done for unfolded strings, although closed strings are addressed. The basic idea of that paper  is that there is an equal probability for a string at any point along its spatial direction. For folded strings, the folding points are special points and it is not unreasonable to think that they play an important role in the process of the breakup.

\item
The process of melting of stringy hadrons was analyzed in \cite{Peeters:2006iu}. It was shown there that at any given temperature there is a critical length and correspondingly critical orbital angular momentum above which the string cannot exist. It is conceivable that folding points along  strings will be more vulnerable when the temperature is raised and thus will modify the sustainability of the corresponding hadrons.
\end{itemize}

\section*{Acknowledgments}
We would like to thank Shimon Yankielowicz for many fruitful conversations and  Simeon Hellerman for a useful discussion. We thank Ori Ganor for his comments on the manuscript. This work was supported in part by a center of excellence supported by the Israel Science Foundation (grant number 2289/18).  DW is supported by the Quantum Gravity Unit at the Okinawa Institute of Science and Technology Graduate University.

\bibliographystyle{JHEP}

\bibliography{clfolnc}

\providecommand{\href}[2]{#2}\begingroup\raggedright\begin{thebibliography}{10}

\bibitem{Hellerman:2013kba}
S.~Hellerman and I.~Swanson, \emph{{String Theory of the Regge Intercept}},
  \href{https://doi.org/10.1103/PhysRevLett.114.111601}{\emph{Phys.Rev.Lett.}
  {\bfseries 114} (2015) 111601}
  [\href{https://arxiv.org/abs/1312.0999}{{\ttfamily 1312.0999}}].

\bibitem{Polchinski:1991ax}
J.~Polchinski and A.~Strominger, \emph{{Effective string theory}},
  \href{https://doi.org/10.1103/PhysRevLett.67.1681}{\emph{Phys.Rev.Lett.}
  {\bfseries 67} (1991) 1681}.

\bibitem{Gross:1993hu}
D.J.~Gross and W.~Taylor, \emph{{Two-dimensional QCD is a string theory}},
  \href{https://doi.org/10.1016/0550-3213(93)90403-C}{\emph{Nucl.\ Phys.\ B}
  {\bfseries 400} (1993) 181}
  [\href{https://arxiv.org/abs/hep-th/9301068}{{\ttfamily hep-th/9301068}}].

\bibitem{Ganor:1994rm}
O.~Ganor, J.~Sonnenschein and S.~Yankielowicz, \emph{{Folds in 2-D string
  theories}}, \href{https://doi.org/10.1016/0550-3213(94)90275-5}{\emph{Nucl.
  Phys.} {\bfseries B427} (1994) 203}
  [\href{https://arxiv.org/abs/hep-th/9404149}{{\ttfamily hep-th/9404149}}].

\bibitem{Pawelczyk:1994ex}
J.~Pawelczyk, \emph{{Toward QCD string: No folds}},
  \href{https://doi.org/10.1103/PhysRevLett.74.3924}{\emph{Phys. Rev. Lett.}
  {\bfseries 74} (1995) 3924}
  [\href{https://arxiv.org/abs/hep-th/9403175}{{\ttfamily hep-th/9403175}}].

\bibitem{Georgiou:2010an}
G.~Georgiou, \emph{{Two and three-point correlators of operators dual to folded
  string solutions at strong coupling}},
  \href{https://doi.org/10.1007/JHEP02(2011)046}{\emph{JHEP} {\bfseries 02}
  (2011) 046} [\href{https://arxiv.org/abs/1011.5181}{{\ttfamily 1011.5181}}].

\bibitem{Bars:1994xi}
I.~Bars, \emph{{Folded strings in curved space-time}},
  \href{https://arxiv.org/abs/hep-th/9411078}{{\ttfamily hep-th/9411078}}.

\bibitem{Bars:1994sv}
I.~Bars and J.~Schulze, \emph{{Folded strings falling into a black hole}},
  \href{https://doi.org/10.1103/PhysRevD.51.1854}{\emph{Phys. Rev.} {\bfseries
  D51} (1995) 1854} [\href{https://arxiv.org/abs/hep-th/9405156}{{\ttfamily
  hep-th/9405156}}].

\bibitem{Attali:2018goq}
K.~Attali and N.~Itzhaki, \emph{{The Averaged Null Energy Condition and the
  Black Hole Interior in String Theory}},
  \href{https://doi.org/10.1016/j.nuclphysb.2019.114631}{\emph{Nucl. Phys.}
  {\bfseries B943} (2019) 114631}
  [\href{https://arxiv.org/abs/1811.12117}{{\ttfamily 1811.12117}}].

\bibitem{Giveon:2019gfk}
A.~Giveon and N.~Itzhaki, \emph{{Stringy Black Hole Interiors}},
  \href{https://doi.org/10.1007/JHEP11(2019)014}{\emph{JHEP} {\bfseries 11}
  (2019) 014} [\href{https://arxiv.org/abs/1908.05000}{{\ttfamily
  1908.05000}}].

\bibitem{Maldacena:2005hi}
J.M.~Maldacena, \emph{{Long strings in two dimensional string theory and
  non-singlets in the matrix model}},
  \href{https://doi.org/10.1088/1126-6708/2005/09/078,
  10.1142/S0219887806001053}{\emph{JHEP} {\bfseries 09} (2005) 078}
  [\href{https://arxiv.org/abs/hep-th/0503112}{{\ttfamily hep-th/0503112}}].

\bibitem{Giombi:2010bj}
S.~Giombi, R.~Ricci, R.~Roiban and A.A.~Tseytlin, \emph{{Quantum dispersion
  relations for excitations of long folded spinning superstring in $AdS_5 x
  S^5$}}, \href{https://doi.org/10.1007/JHEP01(2011)128}{\emph{JHEP} {\bfseries
  01} (2011) 128} [\href{https://arxiv.org/abs/1011.2755}{{\ttfamily
  1011.2755}}].

\bibitem{Sonnenschein:2015zaa}
J.~Sonnenschein and D.~Weissman, \emph{{Glueballs as rotating folded closed
  strings}}, \href{https://doi.org/10.1007/JHEP12(2015)011}{\emph{JHEP}
  {\bfseries 12} (2015) 011}
  [\href{https://arxiv.org/abs/1507.01604}{{\ttfamily 1507.01604}}].

\bibitem{Sonnenschein:2019bca}
J.~Sonnenschein, D.~Weissman and S.~Yankielowicz, \emph{{The scattering
  amplitude of stringy hadrons I: Strings with opposite charges on their
  endpoints}},  \href{https://arxiv.org/abs/1906.00976}{{\ttfamily
  1906.00976}}.

\bibitem{Ando:2010nm}
M.~Ando and E.~Sharpe, \emph{{Two-dimensional topological field theories as
  taffy}}, \href{https://doi.org/10.4310/ATMP.2011.v15.n1.a6}{\emph{Adv. Theor.
  Math. Phys.} {\bfseries 15} (2011) 179}
  [\href{https://arxiv.org/abs/1011.0100}{{\ttfamily 1011.0100}}].

\bibitem{Aharony:2010cx}
O.~Aharony and M.~Field, \emph{{On the effective theory of long open strings}},
  \href{https://doi.org/10.1007/JHEP01(2011)065}{\emph{JHEP} {\bfseries 01}
  (2011) 065} [\href{https://arxiv.org/abs/1008.2636}{{\ttfamily 1008.2636}}].

\bibitem{Aharony:2013ipa}
O.~Aharony and Z.~Komargodski, \emph{{The Effective Theory of Long Strings}},
  \href{https://doi.org/10.1007/JHEP05(2013)118}{\emph{JHEP} {\bfseries 05}
  (2013) 118} [\href{https://arxiv.org/abs/1302.6257}{{\ttfamily 1302.6257}}].

\bibitem{Dubovsky:2012sh}
S.~Dubovsky, R.~Flauger and V.~Gorbenko, \emph{{Effective String Theory
  Revisited}}, \href{https://doi.org/10.1007/JHEP09(2012)044}{\emph{JHEP}
  {\bfseries 09} (2012) 044} [\href{https://arxiv.org/abs/1203.1054}{{\ttfamily
  1203.1054}}].

\bibitem{Hellerman:2014cba}
S.~Hellerman, S.~Maeda, J.~Maltz and I.~Swanson, \emph{{Effective String Theory
  Simplified}}, \href{https://doi.org/10.1007/JHEP09(2014)183}{\emph{JHEP}
  {\bfseries 1409} (2014) 183}
  [\href{https://arxiv.org/abs/1405.6197}{{\ttfamily 1405.6197}}].

\bibitem{Sonnenschein:2016pim}
J.~Sonnenschein, \emph{{Holography Inspired Stringy Hadrons}},
  \href{https://doi.org/10.1016/j.ppnp.2016.06.005}{\emph{Prog. Part. Nucl.
  Phys.} {\bfseries 92} (2017) 1}
  [\href{https://arxiv.org/abs/1602.00704}{{\ttfamily 1602.00704}}].

\bibitem{Gubser:2002tv}
S.~Gubser, I.~Klebanov and A.M.~Polyakov, \emph{{A Semiclassical limit of the
  gauge / string correspondence}},
  \href{https://doi.org/10.1016/S0550-3213(02)00373-5}{\emph{Nucl. Phys. B}
  {\bfseries 636} (2002) 99}
  [\href{https://arxiv.org/abs/hep-th/0204051}{{\ttfamily hep-th/0204051}}].

\bibitem{Caron-Huot:2016icg}
S.~Caron-Huot, Z.~Komargodski, A.~Sever and A.~Zhiboedov, \emph{{Strings from
  Massive Higher Spins: The Asymptotic Uniqueness of the Veneziano Amplitude}},
   \href{https://arxiv.org/abs/1607.04253}{{\ttfamily 1607.04253}}.

\bibitem{Sever:2017ylk}
A.~Sever and A.~Zhiboedov, \emph{{On Fine Structure of Strings: The Universal
  Correction to the Veneziano Amplitude}},
  \href{https://doi.org/10.1007/JHEP06(2018)054}{\emph{JHEP} {\bfseries 06}
  (2018) 054} [\href{https://arxiv.org/abs/1707.05270}{{\ttfamily
  1707.05270}}].

\bibitem{Sonnenschein:2018aqf}
J.~Sonnenschein and D.~Weissman, \emph{{Quantizing the rotating string with
  massive endpoints}},
  \href{https://doi.org/10.1007/JHEP06(2018)148}{\emph{JHEP} {\bfseries 06}
  (2018) 148} [\href{https://arxiv.org/abs/1801.00798}{{\ttfamily
  1801.00798}}].

\bibitem{PandoZayas:2003yb}
L.A.~Pando~Zayas, J.~Sonnenschein and D.~Vaman, \emph{{Regge trajectories
  revisited in the gauge / string correspondence}},
  \href{https://doi.org/10.1016/j.nuclphysb.2003.12.006}{\emph{Nucl.Phys.}
  {\bfseries B682} (2004) 3}
  [\href{https://arxiv.org/abs/hep-th/0311190}{{\ttfamily hep-th/0311190}}].

\bibitem{Sonnenschein:2018fph}
J.~Sonnenschein and D.~Weissman, \emph{{Excited mesons, baryons, glueballs and
  tetraquarks: Predictions of the Holography Inspired Stringy Hadron model}},
  \href{https://doi.org/10.1140/epjc/s10052-019-6828-y}{\emph{Eur. Phys. J. C}
  {\bfseries 79} (2019) 326}
  [\href{https://arxiv.org/abs/1812.01619}{{\ttfamily 1812.01619}}].

\bibitem{Lambiase:1995st}
G.~Lambiase and V.V.~Nesterenko, \emph{{Quark mass correction to the string
  potential}}, \href{https://doi.org/10.1103/PhysRevD.54.6387}{\emph{Phys.
  Rev.} {\bfseries D54} (1996) 6387}
  [\href{https://arxiv.org/abs/hep-th/9510221}{{\ttfamily hep-th/9510221}}].

\bibitem{Zahn:2016hxw}
M.~Kozo\v~n and J.~Zahn, \emph{{Semiclassical energy of closed Nambu-Goto
  strings}}, \href{https://doi.org/10.1103/PhysRevD.100.106005}{\emph{Phys.
  Rev. D} {\bfseries 100} (2019) 106005}
  [\href{https://arxiv.org/abs/1610.02813}{{\ttfamily 1610.02813}}].

\bibitem{Abouelsaood:1986gd}
A.~Abouelsaood, C.G.~Callan, Jr., C.R.~Nappi and S.A.~Yost, \emph{{Open Strings
  in Background Gauge Fields}},
  \href{https://doi.org/10.1016/0550-3213(87)90164-7}{\emph{Nucl. Phys.}
  {\bfseries B280} (1987) 599}.

\bibitem{Beccaria:2010ry}
M.~Beccaria, G.V.~Dunne, V.~Forini, M.~Pawellek and A.A.~Tseytlin, \emph{{Exact
  computation of one-loop correction to energy of spinning folded string in
  AdS5 x S5}}, \href{https://doi.org/10.1088/1751-8113/43/16/165402}{\emph{J.
  Phys.} {\bfseries A43} (2010) 165402}
  [\href{https://arxiv.org/abs/1001.4018}{{\ttfamily 1001.4018}}].

\bibitem{Kinar:1998vq}
Y.~Kinar, E.~Schreiber and J.~Sonnenschein, \emph{{Q anti-Q potential from
  strings in curved space-time: Classical results}},
  \href{https://doi.org/10.1016/S0550-3213(99)00652-5}{\emph{Nucl. Phys.}
  {\bfseries B566} (2000) 103}
  [\href{https://arxiv.org/abs/hep-th/9811192}{{\ttfamily hep-th/9811192}}].

\bibitem{Polyakov:1981rd}
A.M.~Polyakov, \emph{{Quantum Geometry of Bosonic Strings}},
  \href{https://doi.org/10.1016/0370-2693(81)90743-7}{\emph{Phys. Lett.}
  {\bfseries B103} (1981) 207}.

\bibitem{Sonnenschein:2017ylo}
J.~Sonnenschein and D.~Weissman, \emph{{The decay width of stringy hadrons}},
  \href{https://doi.org/10.1016/j.nuclphysb.2017.12.017}{\emph{Nucl. Phys.}
  {\bfseries B927} (2018) 368}
  [\href{https://arxiv.org/abs/1705.10329}{{\ttfamily 1705.10329}}].

\bibitem{Peeters:2006iu}
K.~Peeters, J.~Sonnenschein and M.~Zamaklar, \emph{{Holographic melting and
  related properties of mesons in a quark gluon plasma}},
  \href{https://doi.org/10.1103/PhysRevD.74.106008}{\emph{Phys. Rev.}
  {\bfseries D74} (2006) 106008}
  [\href{https://arxiv.org/abs/hep-th/0606195}{{\ttfamily hep-th/0606195}}].

\end{thebibliography}\endgroup

\clearpage
\appendix

\section{Contour integral, explicit calculations} \label{app:contour}
We do the calculation of \(\sum_{n=1}^\infty n = \zeta(-1) = -\frac{1}{12}\) in the contour integral approach. 

This is the explicit calculation whose results were presented in section \ref{sec:ct}. We start from the function
\be f(z) = \sin(\delta z) \ee
with \(0<\delta\leq\pi\), and we calculate explicitly
\be \frac{\pi}{\delta}\sum_{n=1}^\infty n = \frac{1}{2\pi i}\oint dz z \frac{f'(z)}{f(z)} \label{eq:appA} \ee
Using a cutoff \(N\), the integral over the semicircular contour of figure \ref{fig:contour} breaks up into two parts, an integral over the imaginary axis \(I_{im}\) and an integral over a semicircle of radius \(N\), \(I_{sc}\). The integral over the imaginary axis is
\be I_{im} = -\frac{\delta}{2\pi} \int_{-N}^{N}\!\! dy\, y \coth(\delta y) = \frac{\delta }{2\pi}N^2+\frac{\pi}{12\delta}- i N -\frac{1}{\pi} N \log(e^{2\delta N}-1) - \frac{1}{2\pi\delta}\mathrm{Li}_2(e^{2\delta N})\ee
At large \(N\), one can use the expansion
\be \mathrm{Li}_2(\frac{1}{x}) = \frac{\pi^2}{3}+i\pi \log x -\frac12 (\log x)^2 + \mathcal O(x) \ee
then
\be I_{im} = -\frac{\delta}{2\pi} N^2 - \frac{\pi}{12\delta} + \mathcal O(e^{-2\pi N})\ee

The semicircle integral is harder to evaluate. It is
\be I_{sc} = \frac{\delta}{2\pi}N^2 \int_{-\frac\pi 2}^{\frac\pi 2}d\theta e^{2i\theta} \cot(e^{i\theta}\delta N) \ee
One can get the result at large \(N\) by breaking up \(I_{sc}\) into its real and imaginary parts. The imaginary part gives a symmetric integral over an odd function, so it must vanish. The real part is an even function, and we can write
\be I_{sc} = \frac{\delta N^2}{2\pi}\int_0^{\frac\pi2}d\theta\frac{\sin (2 \theta ) \left(1-e^{-4
   \delta  N \sin\theta}\right)+2 \cos (2\theta ) e^{-2 \delta  N \sin\theta} \sin(2 \delta  N \cos \theta)}
   {1+e^{-4 \delta  N \sin\theta}-2 e^{-2\delta  N \sin\theta} \cos (2\delta  N \cos\theta)} \ee
At the \(N\to\infty\) limit, we can discard all the exponents and we are left only with
\be I_{sc} = \frac{\delta N^2}{2\pi}\int_{0}^{\frac\pi2}d\theta \sin(2\theta) = \frac{\pi}{\delta} N^2 \ee
such that
\be \sum_{n=1}^N {\omega_n} = I_{im} + I_{sc} \to \frac{\pi}{2\delta}N^2 - \frac{\pi}{12\delta} \ee
We get the expected quadratic divergence and the finite piece, which is exactly \(\frac{\pi}{\delta}\zeta(-1)\). The divergent piece is subtracted as explained in section \ref{sec:ct}.

In several places in the paper, we wrote that the divergent part can be written in integral form by considering the contour integral over an asymptotic form of \(f(z)\). In all cases we analyzed in the paper \(I_{sc}\) contributes only divergent terms as we take the cutoff to infinity, and can be dropped from the final expression for the finite intercept. We confirm that we do not subtract any finite piece by checking numerically that \(\frac{I_{sc}}{N^2}-\frac{\pi}\delta\to 0\), which we can also do in the more complicated cases where \(I_{sc}\) is a function of the folding point masses or magnetic field.

We can isolate the divergent part of \(I_{im}\) by taking the asymptotic form of \(f(iy)\) at large \(|y|\),
\be f_A(y)  = \begin{cases} \frac12 i e^{\delta y} & y>0 \\ -\frac12 i e^{-\delta y} & y< 0\end{cases} \ee
By performing the integral on the imaginary axis on this \(f^A(y)\), we get just the quadratic divergence,
\be I_{im}^{div} = -\frac{1}{2\pi i}\int_{-N}^N\!\! dy\, y \frac{f_A^\prime(y)}{f_A(y)} = -\frac{\delta}{\pi}\int_0^N y = -\frac{\delta}{2\pi}N^2 \ee
The finite result is obtained by taking the difference between the full \(I_{im}\) and \(I_{im}^{div}\). There is no need to evaluate \(I_{sc}\) except to show that it is divergent. We usually integrate by parts to write the integral in eq. \ref{eq:appA} as an integral over \(\log f(z)\). Then we can write the finite renormalized sum simply as
\be \sum_{n=1}^\infty \omega_n = \frac{1}{2\pi} \int_{-\infty}^\infty\!\! dy\, \log \frac{f(iy)}{f_A(y)} \ee
In the example discussed here this formula leads to
\be \sum_{n=1}^\infty \omega_n = \frac{1}{\pi} \int_0^\infty \!\! dy\, \log(1-e^{-2 \delta y}) = -\frac{\pi}{12\delta} = \frac\pi\delta \zeta(-1) \ee
In appendix \ref{app:Zeta} we write more examples of such equations between contour integrals made finite by the subtraction described above and the results of Zeta function regularization.

\section{Comparison of Zeta function regularization and contour integral subtraction} \label{app:Zeta}
In sections \ref{sec:quantc} and \ref{sec:quanto} we computed the corrections to the intercept with a finite mass term present at the folds. We can perform the regularization of the divergent sum over the eigenfrequencies in two ways. The first method is to write an approximate solution of the eigenfrequency equation for small masses, and then perform a Zeta function regularization order by order. The second method is to convert the infinite sum into a contour integral, and subtract the divergent terms from the result. This method gives a closed expression in integral form, which can be expanded for small masses and compared with the Zeta function result. The two methods give the same result for all examples discussed in this paper.

To do the Zeta function we start from an expansion of the form
\be \omega_n = n + c_1(n)\epsilon + c_2(n) \epsilon^2 + c_3(n) \epsilon^3 + \ldots \ee
where the coefficients \(c_i\) are functions of \(n\). The \(c_i\) in our solutions grow with \(n\), typically as \(n^i\). Therefore at finite \(\epsilon\) we cannot use the expansion for a good approximate solution of \(\omega_n\) for any \(n\). On the other hand, when we write the expansion of the intercept
\be a = -\frac12\sum_{n=1}^\infty \omega_n = a_0 + a_1 \epsilon + a_2 \epsilon^2 + a_3 \epsilon^3 + \ldots \ee
the coefficients will be
\be a_i = -\frac12\sum_{n=1}^\infty c_i(n) \ee
In this sum, the contribution from large \(n\) is regularized by use of the Zeta function, so the intercept unlike \(\omega_n\) does have a good convergent expansion in \(\epsilon\).

If the coefficient \(c_i\) is expressible as a polynomial of \(n\), then the prescription of the Zeta function regularization is to replace each sum \(\sum n^k\) with \(\zeta(-k)\).

The results of the contour integral method are written in the text. There we have a closed expression for the intercept as an integral. The comparison with the Zeta function is made by expanding the integrand in the same small parameter \(\epsilon\) and writing an integral expression for the coefficients \(a_i\).

\subsection{Closed string} \label{app:Zeta_closed}
For the closed strings, as described in section \ref{sec:quantc}, most generally we have two expansion parameters for two folding point masses, \(\epsilon_1 = \frac{1}{\gamma_0}\) and \(\epsilon_2 = \frac{1}{\gamma_\ell}\), which we can vary independently of each other.

For the \textbf{odd transverse} modes on the closed string, the eigenfrequencies are given exactly by
\be \tilde\omega_n = \frac\pi\delta n \ee
The contour integral calculation of 
\be a = -\frac12\sum_{n=1}^{\infty}\omega_n = -\frac{\pi}{2\delta}\zeta(-1) = \frac{\pi}{24\delta} \ee
is what we detailed above in appendix \ref{app:contour}.

The \textbf{even transverse} modes and the \textbf{odd planar mode} both have the spectrum
\be \omega_n = n + \frac{1}{3\pi}(n^3-n)(\epsilon_1^3+\epsilon_2^3) + \ldots \ee
Therefore we can write the expansion of the intercept
\be a = a_0 + a_3(\epsilon_1^3+\epsilon_2^3) + \ldots \ee
where
\be a_0 = -\frac12 \zeta(-1) = \frac{1}{24} \ee
and
\be a_3 = -\frac{1}{2}\sum_{n=1}^\infty \frac{n^3-n}{3\pi} = -\frac{1}{6\pi}(\zeta(-3)-\zeta(-1)) = -\frac{11}{720\pi}
\ee
In the integral form, expanding equation \ref{eq:atcei},
\be a_3 = \frac{1}{6\pi}\int_0^\infty (y^3+y)(1-\coth(\pi y)) \ee

The \textbf{even planar} mode is nearly identical to the previous modes, with the only difference at order \(\epsilon^3\) a factor of \(-\frac12\), so
\be \omega_n = n - \frac{1}{6\pi}(n^3-n)(\epsilon_1^3+\epsilon_2^3) + \ldots \ee
Similarly to the above,
\be a_0 = -\frac12\zeta(-1) \qquad a_3 = \frac{1}{12\pi}(\zeta(-3)-\zeta(-1)) = \frac{11}{1440\pi} \ee
And expanding \ref{eq:acpei},
\be a_3 = -\frac{1}{12\pi}\int_0^\infty (y^3+y)(1-\coth(\pi y)) \ee

\subsection{Open string} \label{app:Zeta_open}
\subsubsection{Transverse modes}
The order by order solution of eq. \ref{eq:wto} for \(\omega_n\) is
\be \omega_n = n + c_1\epsilon + c_2\epsilon^2 + c_3\epsilon_3 + \ldots \label{eq:wexpz} \ee
with \(\epsilon = \frac{1}{\gamma_f}\), and the coefficients
\begin{align}
c_1 &= \frac1\pi n(1-\cos(2n\phi)) \\
c_2 &= \frac{4}{\pi^2} n \sin^3(n\phi)\left(\sin(n\phi)-(\pi-2\phi)n\cos(n\phi)\right) \\
c_3 &= -\frac{1}{3\pi}n(1+\cos(2n\phi)) + \frac{4}{\pi^3}n \sin^4(n\phi)(1-\cos(2n\phi)) - \frac{24}{\pi^3}n^2(\pi-2\phi)\sin^5(n\phi)\cos(n\phi) + \nonumber \\
&\qquad+ \frac{2}{3\pi}n^3 + \frac{2}{\pi^3}(\pi-2\phi)^2\sin^4(n\phi)(1+3\cos(2n\phi))-\frac{8}{\pi^3}n^3(\pi-\phi)\phi \sin^4(n\phi) + \frac{4}{3\pi}n^3 \sin^6(n\phi)
\end{align}
 
 We can use the Zeta function regularization to compute order by order the coefficients in the sum
 \be a_t = -\frac12\sum_{n=1}^\infty \omega_n = a_0 + a_1 \epsilon + a_2 \epsilon^2 + a_3 \epsilon^3 + \ldots \ee
The zeroth order is the usual result of
\be a_0 = -\frac12\sum_{n=1}^\infty n = -\frac12\zeta(-1) = \frac{1}{24} \ee
To compute the other coefficients we write them as an expansion in \(\phi\) such that we get only powers of \(n\) in the sum. Then we replace each term involving a sum over \(n\) with the appropriate Zeta function. In this way we obtain
\begin{align} a_1 &= -\frac12\sum_{n=1}^\infty c_1 = -\frac{1}{2\pi}\sum_{n=1}^\infty n(1-\cos(2n\phi)) \nonumber \\
&= -\frac{1}{2\pi}\sum_{n=1}^\infty \sum_{k=1}^\infty \frac{(-1)^{k+1}}{(2k)!} (2\phi)^{2k} n^{2k+1} \nonumber \\
&= -\frac{1}{2\pi}\sum_{k=1}^\infty \frac{(-1)^{k+1}}{(2k)!} (2\phi)^{2k} \zeta(-1-2k) \nonumber \\
&= \frac{1}{2\pi}\sum_{k=1}^\infty \frac{(-1)^{k+1}}{(2k)!(2k+2)} B_{2k+2}(2\phi)^{2k} \nonumber \\
&= \frac{1}{24\pi}\left(1+\frac{3}{\phi^2}-\frac{3}{\sin^2\phi}\right)
\end{align}
Between the third and fourth line we used the identity
\be B_{n} = (-1)^{n+1} n \zeta(1-n) \qquad \Rightarrow \qquad \zeta(-1-2k) = -\frac{B_{2k+2}}{2k+2} \ee
where \(B_n\) are the Bernoulli numbers. This is not a necessary step, but the \(B_n\) are known to appear in the expansions of the tangent, cotangent, and cosecant functions, and when the sum is written in terms of Bernoulli numbers rather than Zeta functions it is recognizable to Wolfram Mathematica's algorithms.

The exact result for \(a_t = -\frac{1}{2}\sum{\omega_n}\) in its integral form is given in eq. \ref{eq:a_t_open_phi}. We expand the integrand in \(\epsilon = \frac{1}{\gamma_f}\) to give an integral expression for the coefficient \(a_1\), which matches exactly the Zeta function result, it is
\be a_1 = -\frac{2}{\pi}\int_0^\infty dy \frac{\sinh^2(\phi y) y}{e^{2\pi y}-1} \ee

For the next coefficient it is simpler to separate into even and odd parts, \(c_2 = c_2^e + c_2^o\),
\begin{align}
c_2^o &= \frac{1}{2\pi}n^2 \left[ \sin (4 n \phi )-2\sin (2 n \phi )\right] \\
c_2^e &= \frac{1}{2\pi^2}n \left[\cos(4n\phi)-4\cos(2n\phi)+3\right]-\frac{1}{\pi^2}n^2 \phi \left[\sin(4n\phi)-2\sin(2n\phi)\right]
\end{align}
So when expanding in \(\phi\) \(c_2^o\) contains only odd powers of \(\phi\), and \(c_2^e\) only even powers.
Now we compute \(a_2\) in the same way we computed \(a_1\), except that we do even and odd parts separately for simplicity. The results are
\begin{align} a_2^o &= -\frac12\sum_{n=1}^\infty c_2^o = -\frac{1}{2\pi}\sum_{n=1}^\infty \sum_{k=1}^\infty \frac{(-1)^{k}}{(2k+1)!}(4^k-1)n^{2k+3}(2\phi)^{2k+1} \nonumber \\
 &=  -\frac{1}{2\pi}\sum_{k=1}^\infty \frac{(-1)^k}{(2k+1)!}(4^k-1)\zeta(-3-2k)(2\phi)^{2k+1} \nonumber \\ 
 &= \frac{1}{2\pi}\sum_{k=1}^\infty \frac{(-1)^k}{(2k+1)!(2k+4)}(4^k-1)B_{2k+4}(2\phi)^{2k+1} \nonumber \\ 
 &= \frac{1}{128\pi}\left(\frac{15}{\phi^3} -\frac{15\cos\phi}{\sin^3\phi}-\frac{\sin\phi}{\cos^3\phi}\right) \\ \nonumber \\
a_2^e &= -\frac12\sum_{n=1}^\infty c_2^e = -\frac{1}{4\pi^2}\sum_{n=1}^\infty \sum_{k=1}^\infty \frac{(-1)^k}{(2k)!}(4^k-4)(k+1)n^{2k+1}(2\phi)^{2k} \nonumber \\ &= -\frac{1}{4\pi^2} \sum_{k=1}^\infty \frac{(-1)^k}{(2k)!}(4^k-4)(k+1)\zeta(-1-2k)(2\phi)^{2k} \nonumber \\ &= \frac{1}{4\pi^2}\sum_{k=1}^\infty \frac{(-1)^k}{(2k)!(2k+2)}(4^k-4)(k+1)B_{2k+2}(2\phi)^{2k} \nonumber \\
&= \frac{1}{64 \pi ^2}\left(4+\frac{15\phi\cos\phi}{\sin^3\phi}-\frac{15}{\sin^2\phi}+\frac{\phi\sin\phi}{\cos^3\phi}+\frac{1}{\cos^2\phi}\right)
\end{align}
And finally
\be a_2 = a_2^e+a_2^o = \frac{1}{128\pi^2}\left[8+\frac{15 \pi }{\phi ^3}-\frac{30}{\sin^2\phi}+\frac{2}{\cos^\phi}-(\pi-2\phi)\left(\frac{15\cos\phi}{\sin^3\phi}+\frac{\sin\phi}{\cos^3\phi}\right)\right] \ee
which matches with the integral
\be a_2 = \frac{4}{\pi}\int_0^\infty dy \frac{\sinh^3(\phi y)[\cosh(\phi y)-e^{(2\pi-\phi)y}]y^2}{(e^{2\pi y}-1)^2}\ee

At order \(\epsilon^3\), we again split the calculation into its odd and even parts. The expressions are longer, but the method is the same.
\begin{align} a_3^o &= -\frac12\sum_{n=1}^\infty c_3^o = -\frac{1}{4\pi^2}\sum_{n=1}^\infty \sum_{k=1}^\infty \frac{(-1)^{k}}{(2k+1)!}(2^{2k+3}-3^{2k+1}-5)(k+2)n^{2k+3}(2\phi)^{2k+1} \nonumber \\
&= -\frac{1}{4\pi^2}\sum_{k=1}^\infty \frac{(-1)^{k}}{(2k+1)!}(2^{2k+3}-3^{2k+1}-5)(k+2)\zeta(-3-2k)(2\phi)^{2k+1} \nonumber \\
&= \frac{1}{8\pi^2}\sum_{k=1}^\infty \frac{(-1)^{k}}{(2k+1)!}(2^{2k+3}-3^{2k+1}-5)B_{2k+4}(2\phi)^{2k+1} \nonumber \\
&= -\frac{1}{32\pi^2}\bigg(\frac{15\cos\phi}{\sin^3\phi}-\frac{12\cos(2\phi)}{\sin^3(2\phi)}+\frac{3\cos(3\phi)}{\sin^3(3\phi)}+\frac{\phi}{\sin^4\phi \cos^4\phi}+ \nonumber \\ &\qquad\qquad\qquad\qquad\qquad +\frac{\phi\cos(4\phi)}{2\cos^4\phi\sin^4\phi} - \frac{5\phi(1+2\cos^2\phi)}{\sin^4\phi}-\frac{3\phi(2+\cos(6\phi)}{\sin^4(3\phi)} \bigg)
\end{align}

\begin{align} a_3^e &= \nonumber \frac{\zeta(-1)-\zeta(-3)}{3\pi}+ \nonumber \\
&\,\,+\frac{1}{24\pi^3}\sum_{k=1}^\infty \frac{(-1)^k}{(2 k)!} \bigg(\left(\left(15-3\ 2^{2 k+1}+3^{2 k}\right) k (2 k+5)-9\ 2^{2 k+1}+3^{2 k+1}+4 \pi ^2+45\right) \zeta (-2k-1) - \nonumber \\
&\qquad\qquad\qquad -4 \pi ^2 \left(3-3\ 2^{2 k}+3^{2 k}\right) \zeta (-2 k-3)\bigg) (2\phi)^{2 k} = \nonumber \\
&= \frac{1}{103680\pi^3}\bigg(\frac{18305 \pi ^2}{\phi ^4}+\frac{4320 \pi ^2}{\phi ^2}-\frac{4860 \phi ^2 \sin ^2(6 \phi )}{\sin (3 \phi )^6}-\frac{9720\phi ^2}{\sin (3 \phi )^4}-\frac{3240 \left(5 \phi ^2+2 \pi ^2\right)}{\sin (\phi )^4} +\nonumber \\ &\,\,\,+\frac{6480 \left(4 \phi ^2+\pi^2\right)}{\sin (2 \phi )^4}-\frac{1080 \left(6 \left(5 \phi ^2+2 \pi ^2\right) \cot ^2(\phi )-90 \phi  \cot (\phi )+4
   \pi ^2+45\right)}{\sin (\phi )^2}+\nonumber \\ &\,\,\,+\frac{6480 \left(8 \phi ^2 \cot ^2(2 \phi )-12 \phi  \cot (2 \phi )+3\right)}{\sin (2\phi )^2}+\frac{3240 \pi ^2 \sin ^2(4 \phi )}{\sin (2 \phi )^6}+\frac{9720 \phi  \sin (6 \phi )}{\sin (3 \phi)^4}-\nonumber \\ &\,\,\,-\frac{3240}{\sin (3 \phi )^2}-\frac{2160 \pi ^2}{\sin (3 \phi )^4}-\frac{4320 \pi ^2 \cot ^2(3 \phi )}{\sin (3 \phi)^2}-1584 \pi ^2+10800\bigg)
 \end{align}
These expressions match with
\begin{align} a_3 &= a_3^e+a_3^o = \frac{1}{6\pi}\int_0^\infty dy \bigg(\frac{y (3 \cosh (2 y \phi )+1)}{1-e^{2 \pi 
   y}}+ \nonumber \\
   &\qquad + \frac{y^3 e^{4 \pi  y} (3 e^{-6 y \phi }-12 e^{-4 y \phi }+18 e^{-2 y \phi }+3 e^{2 y \phi }-8)}{\left(1-e^{2 \pi  y}\right)^3}\nonumber \\
     &\qquad+ \frac{y^3 e^{2 \pi  y}
   (-3 e^{-6 y \phi }+9 e^{-4 y \phi }-6 e^{-2 y \phi }+9 e^{2 y \phi }-3 e^{4 y \phi }-14)}{\left(1-e^{2 \pi  y}\right)^3}\nonumber \\
   &\qquad+ \frac{y^3 e^{2 \pi  y}(6 \cosh (2 y \phi )-6
   \cosh (4 y \phi )+2 \cosh (6 y \phi )+2)}{\left(1-e^{2 \pi  y}\right)^3}\bigg)
   \end{align}
   
\subsubsection{Planar mode}
The order by order solution of eq. \ref{eq:wpo} for \(\omega_n\) is
\be \omega_n = n +  \frac{1-3\cos(2n\phi)}{6\pi}(n^3-n)\epsilon^3 + \ldots \ee
The coefficient of the \(\epsilon^3\), leading order correction, is
\begin{align} a_3 &= -\frac{1}{12\pi}\sum_{n=1}^\infty (n^3-n)(1-3\cos(2n\phi)) \nonumber \\
&=  \frac{1}{12\pi}\sum_{n=1}^\infty \left(2(n^3-n) + 3\sum_{k=1}^\infty \frac{(-1)^k}{(2k)!}(n^{2k+3}-n^{2k+1})(2\phi)^{2k}\right) \nonumber \\
&=  \frac{1}{12\pi}\left(2(\zeta(-3)-\zeta(-1)) + 3\sum_{k=1}^\infty \frac{(-1)^k}{(2k)!}(\zeta(-3-2k)-\zeta(-1-2k))(2\phi)^{2k}\right) \nonumber \\ 
&= \frac{1}{12\pi}\left[2(\zeta(-3)-\zeta(-1)) + 3\sum_{k=1}^\infty \frac{(-1)^k}{(2k)!}\left(\frac{B_{2k+2}}{2k+2}-\frac{B_{2k+4}}{2k+4}\right)(2\phi)^{2k}\right] \nonumber \\ 
&= -\frac{1}{24\pi}\left(\frac{9}{4 \phi ^4}-\frac{9}{4\sin^4\phi} +\frac{3}{2 \phi ^2}+\frac{11}{60}\right)
\end{align}
This matches exactly with the contour integral result of eq. \ref{eq:apo}, which is
\be a_3 = -\frac{1}{6\pi}\int_0^\infty dy \frac{(y^3+y)(3\cosh(2\phi y)-1)}{1-e^{2\pi y}} \ee

\end{document}